\documentclass[
reprint,
twocolumn,
superscriptaddress,
amsmath,amssymb,
aps,
prx,
floatfix,
]{revtex4-2}

\usepackage{libertine}
\usepackage{graphicx}
\usepackage{dcolumn}
\usepackage{bm}
\usepackage{physics}
\usepackage{bbm}
\usepackage{siunitx}
\usepackage{amsmath}
\usepackage{dsfont}
\usepackage{fontspec}
\usepackage{array,booktabs}
\usepackage{comment} 
\usepackage{xcolor}
\usepackage{rotating} 
\usepackage{textcomp, gensymb} 
 
\usepackage{booktabs}
\usepackage[symbol]{footmisc}
\usepackage{placeins} 
\usepackage{float}
\definecolor{NewBlue}{rgb}{0.0313,0.090,0.208}
\definecolor{NewLightBlue}{rgb}{0,0.404,0.58}
\definecolor{NewRed}{rgb}{0.573,0.082,0.082}
\definecolor{NewBlack}{rgb}{0.,0.0,0.0}
\definecolor{NewBlue}{rgb}{0, 0, 0.41}
\definecolor{LWBlue}{rgb}{0.000, 0.200, 0.290}
\usepackage{cancel}
\usepackage[normalem]{ulem}
\usepackage[version=4]{mhchem}

\newcommand{\vsi}{\textrm{V}_\textrm{Si}^\textrm{-}}
\newcommand{\vtwo}{\mathrm{V2}}
\newcommand{\gdif}{\gamma_\mathrm{d}}
\newcommand{\gion}{\gamma_\mathrm{i}}

\newcommand{\mc}{m}
\newcommand{\Co}{C_0}
\newcommand{\Cmean}{C}

\newcommand{\subfigref}[2]{\hyperref[#1]{Fig.~\ref*{#1}#2}}
\DeclareSIUnit \vpp {\ensuremath{\mathrm{V_{pp}}}}
\newcommand{\Ttwo}{\ensuremath{T_{2}}}
\newcommand{\Ttwodd}{\ensuremath{T_{2}^\mathrm{DD}}}
\newcommand{\Ttwohahn}{\ensuremath{T_{2}^\mathrm{Hahn}}}
\newcommand{\Ttwostar}{\ensuremath{T_{2}^{*}}}

\definecolor{RED}{named}{red}
\setcitestyle{super,sort&compress}

\usepackage[colorlinks,
    linkcolor=LWBlue,
    citecolor=LWBlue,
    urlcolor=LWBlue]{hyperref}

\DeclareSIUnit{\gauss}{G}

\begin{document}

\title{Laser-induced creation of coherent V2 centers in bulk-grown silicon carbide}
\author{L.J. Feije}
\thanks{These authors contributed equally}
\author{G.M. Timmer}
\thanks{These authors contributed equally}
\author{Y. Hu}
\author{R. Karababa}
\author{G.L. van de Stolpe}
\author{T. Martens}
\author{S.J.H. Loenen}
\author{T.B.A. Durant}
\author{A. Das}

\affiliation{QuTech, Delft University of Technology, PO Box 5046, 2600 GA Delft, The Netherlands}%

\affiliation{Kavli Institute of Nanoscience Delft, Delft University of Technology,
PO Box 5046, 2600 GA Delft, The Netherlands}
\author{A.M. Day}
\author{E.L. Hu}
\affiliation{John A. Paulson School of Engineering and Applied Sciences, Harvard University, Cambridge, MA, USA}
\author{T.H. Taminiau}
\email{t.h.taminiau@tudelft.nl}
\affiliation{QuTech, Delft University of Technology, PO Box 5046, 2600 GA Delft, The Netherlands}%

\affiliation{Kavli Institute of Nanoscience Delft, Delft University of Technology,
PO Box 5046, 2600 GA Delft, The Netherlands}

\date{\today}
\begin{abstract}
\noindent Solid-state spin defects are promising qubits for quantum network nodes. A key challenge towards larger networks is creating defects with high yield into nanophotonic devices, while maintaining good optical and spin properties. Here, we demonstrate the creation of single $\vtwo$ centers in nanopillars fabricated from commercial bulk-grown 4H-silicon carbide using a pulsed above-bandgap (UV) laser. We observe an eleven-fold increase in the $\vtwo$ center occurrence after UV laser illumination. These laser-induced $\vtwo$ centers exhibit narrow optical linewidths and spectral diffusion rates comparable to naturally occurring $\vtwo$ centers in nanopillars of the same material. Furthermore, we measure a spin coherence time of $\Ttwodd = \SI{3.6\pm0.3 }{\milli\second}$ under dynamical decoupling, consistent with dephasing by the nuclear-spin bath. This demonstration of the in-situ, post-fabrication generation of coherent $\vtwo$ centers in nanostructures in widely available bulk-grown 4H-SiC, shows the potential for above-bandgap laser illumination for scalable defect creation in integrated photonic devices. 
\end{abstract}

\maketitle
\section*{Introduction}
\noindent Solid-state spin defects provide promising qubits for quantum networks, as they combine high-fidelity spin-photon entanglement with access to nuclear spins with long coherence times \cite{bradleyTenQubitSolidStateSpin2019,stolkMetropolitanscaleHeraldedEntanglement2024,pompiliRealizationMultinodeQuantum2021}. Their compatibility with integrated photonics offers a scalable path towards large-scale quantum networks \cite{knautEntanglementNanophotonicQuantum2024,lukin4HsiliconcarbideoninsulatorIntegratedQuantum2020, harrisHighFidelityControlStrongly2025}.
A central challenge is the deterministic placement of defects within sub-wavelength nanophotonic structures while preserving good optical and spin coherence \cite{rufOpticallyCoherentNitrogenVacancy2019,vandamOpticalCoherenceDiamond2019,gonzalez-tudelaLightMatterInteractions2024,majetyWaferscaleIntegrationFreestanding2025}.

Silicon carbide (4H-SiC) hosts a variety of optically active spin defects \cite{steidlSingleV2Defect2025,christleIsolatedElectronSpins2015,sonDevelopingSiliconCarbide2020,andersonFivesecondCoherenceSingle2022,majetyWaferscaleIntegrationFreestanding2025,normanSub2KelvinCharacterization2025,hessenauerCavityEnhancementV22025,liuSiliconVacancyCenters2024,astnerVanadiumSiliconCarbide2024,wolfowiczVanadiumSpinQubits2020,cilibrizziUltranarrowInhomogeneousSpectral2023}. The negatively charged single k-site silicon vacancy center ($\vtwo$) is a promising candidate as long spin coherence times and near-lifetime-limited optical transitions, required for remote entanglement generation, have been demonstrated in nanophotonic devices \cite{babinFabricationNanophotonicWaveguide2022,heilerSpectralStabilityV22024,vandestolpeCheckprobeSpectroscopyLifetimelimited2025}. 

Several methods have successfully demonstrated the creation of localised single $\vtwo$ centers, such as masked \ce{He^+} ion implantation \cite{babinFabricationNanophotonicWaveguide2022,wangOnDemandGenerationSingle2019} and focused ion beam (FIB) implantation \cite{pavunnyArraysSiVacancies2021,heMasklessGenerationSingle2023,wangScalableFabricationSingle2017}, including with in-situ monitoring \cite{chandrasekaranHighYieldDeterministicFocused2023}. Pulsed-laser writing is a compelling alternative that provides defect creation with good spatial precision, after device fabrication \cite{chenLaserWritingIndividual2019,dayLaserWritingSpin2023,haoLaserWritingSpin2025,jonesScalableRegistrationSingle2025}. In addition, laser writing is compatible with in-situ monitoring  \cite{chenLaserWritingIndividual2019}, potentially enabling deterministic defect creation. 
Pioneering experiments in 4H-SiC have shown that below-bandgap laser writing can create colour centers in bulk \cite{haoLaserWritingSpin2025,jonesScalableRegistrationSingle2025,chenLaserWritingScalable2019}, while above-bandgap laser writing has further enabled the creation of $\vtwo$ centers in 1D nanophotonic crystal cavities \cite{dayLaserWritingSpin2023}.

However, the optical and spin coherence of single laser-written defects in 4H-SiC, which are essential for many quantum applications, have thus far remained unexplored.

Here, we demonstrate laser-induced creation of $\vtwo$ centers in nanopillars with preserved optical and spin coherence. We use an above-bandgap laser and commercial bulk-grown High-Purity Semi-Insulating (HPSI) 4H-SiC, widely available as 4-6 inch wafers.

We first identify the pulse energy window that enables defect formation without inducing surface amorphisation. Subsequently, we show the creation of single laser-induced $\vtwo$ centers and characterise their optical properties. We show near-lifetime-limited optical linewidths and slow spectral diffusion rates, similar to naturally occurring $\vtwo$ centers in the same type of nanopillars. Finally, we demonstrate long electron-spin coherence times (up to \SI{3.6\pm0.3}{\milli\second}) for laser-induced $\vtwo$ centers, similar to values reported in high-quality epitaxial 4H-SiC\cite{steidlSingleV2Defect2025,kunaLocalizationCoherentControl2025}. 
The demonstration of spatially selective defect creation after device fabrication, with good optical and spin coherence in nano-structures made of commercial bulk-grown wafers, indicates that above-bandgap pulsed-laser illumination is a promising route towards scalable generation of coherent $\vtwo$ centers in 4H-SiC devices.
\section*{Results}
\subsection*{Laser-induced amorphization}
\label{subsec:defect_creation}
\begin{figure*}[ht!]
  \includegraphics[width=1 \textwidth]{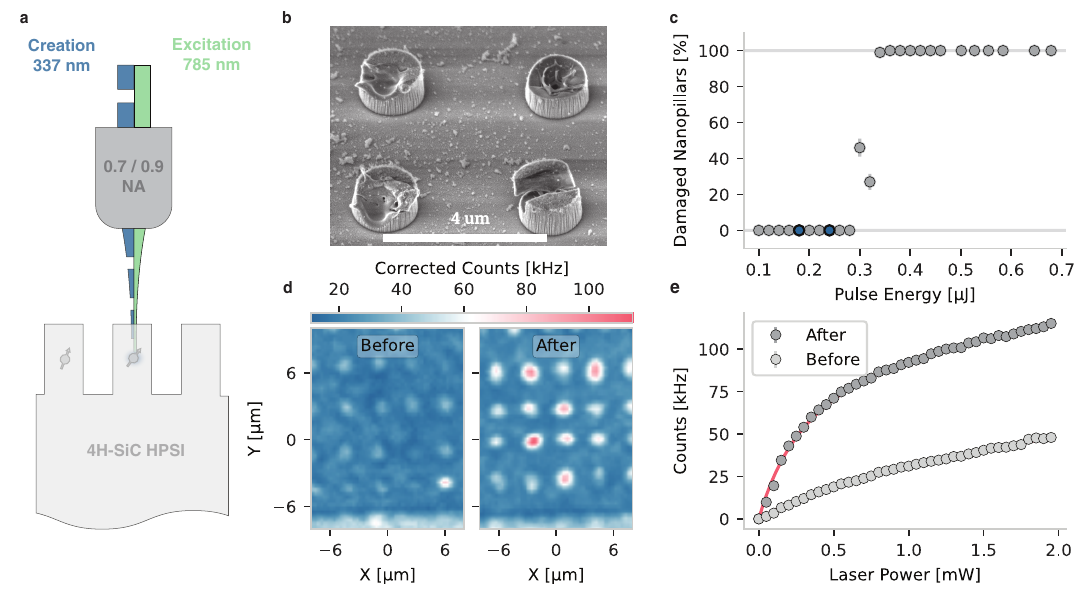}
  \caption{\label{fig:defect_creation} \textbf{Laser-induced defect creation using an above-bandgap pulsed laser.} \textbf{a)} Schematic of the laser-induced creation and detection of defects in HPSI 4H-SiC. A 0.7 NA objective focuses an above-bandgap (\SI{337}{\nano\meter}) pulsed (\SI{3}{\nano\second}) laser onto the center of the nanopillar to create defects. For imaging, we use a 0.9 NA objective and a $\SI{785}{\nano\meter}$ laser for off-resonant excitation. \textbf{b)} Scanning electron microscopy image of the HPSI amorphized nanopillars that were exposed to a single UV pulse with energies higher than the LIAT ($\SI{0.28}{\micro\joule}$). The nanopillars are \raisebox{0.4ex}{\tiny$\sim$}$\SI{1.2}{\micro\meter}$ in diameter and \raisebox{0.4ex}{\tiny$\sim$}$\SI{4}{\micro\meter}$ spaced from center to center. \textbf{c)} Amorphisation statistics of nanopillars using a single pulse. For each pulse energy, one hundred nanopillars were each illuminated with a single pulse, and the percentage of nanopillars that showed visible damage is indicated on the y-axis. Blue dots indicate $\SI{0.18}{\micro\joule}$ and $\SI{0.24}{\micro\joule}$, which are used for \autoref{fig:amound_defects} and \autoref{fig:diffusion_ple}. See \autoref{fig:supp_damage_np_bulk} for SEM pictures of amorphisation. \textbf{d)} 2D PL scans before and after a single UV pulse of $\SI{0.24}{\micro\joule}$ at each pillar. The counts for both measurements were corrected to have the same background counts in bulk close to the nanopillars (before counts $\times 1.524$, and after counts $\times 0.656$, see \ref{subsec:app_before_after_2dpl} for correction method).
  \textbf{e)} Fluorescence saturation measurements before and after a single UV pulse of $\SI{0.24}{\micro\joule}$ on the same pillar. }
\end{figure*}

\noindent We use an above-bandgap ($\SI{337}{\nano\meter}, \SI{3.68}{\electronvolt}$) pulsed ($\SI{3}{\nano\second}$) laser (similar as Day \emph{et al.}\cite{dayLaserWritingSpin2023}) focussed on the sample via a \SI{0.7}{NA} objective and investigate single-pulse defect creation in HPSI 4H-SiC with a bandgap of \SI{3.26}{\electronvolt} \cite{sonChargeStateControl2021} (\subfigref{fig:defect_creation}{a,b}).
Given the high efficiency of electron-hole pair generation at above-bandgap excitation, the material is susceptible to irreversible damage under intense irradiation \cite{dayLaserWritingSpin2023}. To mitigate this, we first determine the laser-induced amorphization threshold (LIAT) \cite{woodLaserInducedDamageOptical2003a,nasiriInvestigationLaserInduced2021} for both nanopillars and bulk HPSI 4H-SiC. This is achieved by exposing one hundred discrete spots per pulse energy. After UV illumination, scanning electron microscopy (SEM) is used to quantify the number of sites exhibiting resolvable surface amorphization (see \ref{subsec:app_damage_threshold}). We note that our analysis is limited to SEM-detectable surface amorphization, which excludes sub-surface lattice damage.

Despite being limited to surface-level observations, our analysis enables the identification of three distinct regimes: no visible amorphization, probabilistic amorphization, and deterministic amorphization, observed in both the nanopillars (\subfigref{fig:defect_creation}{c}) and bulk material (\ref{subsec:app_damage_threshold}). We find that the LIAT for nanopillars  ($>\SI{0.28}{\micro\joule}$) is lower compared to bulk ($>\SI{0.32}{\micro\joule}$), showing that it is dependent on the nanophotonic structure (consistent with previous studies \cite{oliveiraHighaspectratioUltratallSilica2025}). 

\subsection*{Laser-induced defect creation}
\noindent With the LIAT determined, we verify that defect creation is possible at energies below this threshold. We perform off-resonant (785 nm) 2D photoluminescence (PL) scans before and after a single UV laser pulse for pulse energies below the LIAT, employing a \SI{0.9}{NA} objective to improve collection efficiency (\ref{subsec:app_collection_efficiency}). An example of one of these 2D PL scans can be found in \subfigref{fig:defect_creation}{d}. 
For pulse energies \SI{\geq0.16}{\micro\joule}, the 2D PL scans show an increase in PL for nearly all nanopillars (see \ref{subsec:app_before_after_2dpl}), indicating sub-surface lattice deformation or laser-induced defect generation \cite{liuColorCentersCrystal2025}.
Next, we perform fluorescence saturation measurements under off-resonant laser excitation on each individual nanopillar. These measurements reveal early saturation effects, associated with quantum-defect emission, for UV pulse energies $\geq\SI{0.16}{\micro\joule}$, see \ref{subsec:app_rt_violin}.
This suggests that laser-induced defect creation occurs over a broad parameter space, providing a usable energy window (\SIrange{0.16}{0.28}{\micro\joule}) in which colour centers can be introduced without inducing amorphisation.  

To further investigate the number of defects generated by a single UV pulse, we perform fluorescence saturation measurements on an additional 64 nanopillars before and after a single \SI{0.18}{\micro\joule} UV pulse (low-probability regime), and on 64 nanopillars exposed to a \SI{0.24}{\micro\joule} pulse (higher-probability regime). An example of a nanopillar fluorescence saturation measurement before and after a single pulse can be seen in \subfigref{fig:defect_creation}{e}. 

\begin{figure*}[ht!]
  \centering
  \includegraphics[width=1 \textwidth]{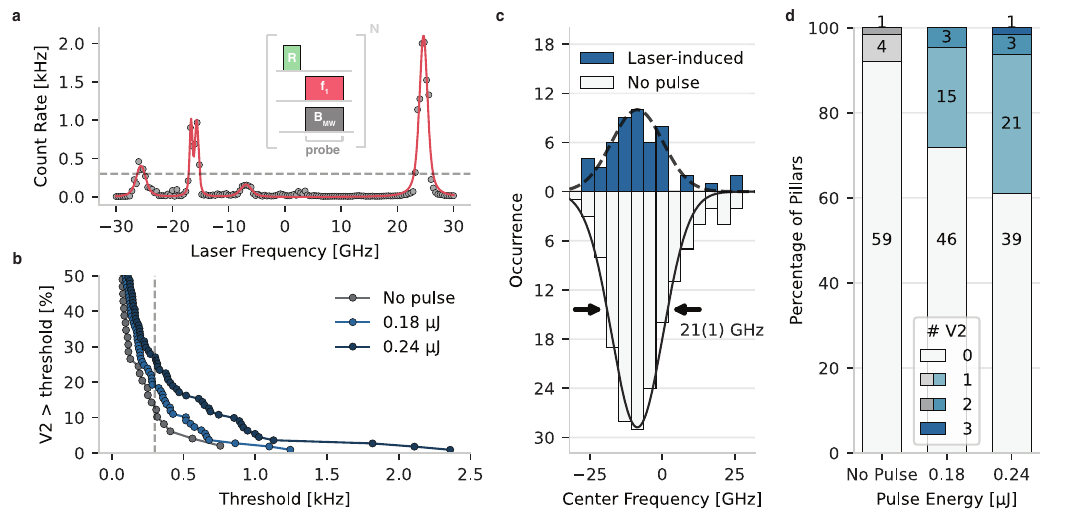}
  \caption{\label{fig:amound_defects} \textbf{Photoluminescence Excitation spectroscopy (PLE) and number of $\vtwo$ centers.} \textbf{a)} Single PLE of a nanopillar with three $\vtwo$ centers passing a threshold (dashed grey line) at \SI{0.3}{\kilo\hertz}. Laser frequency is offset from \SI{327.112}{\tera\hertz}. Inset shows the experimental sequence, green indicates an off-resonant (repump) pulse (\SI{10}{\micro\second}, \SI{10}{\micro\watt}), red indicates a single resonant laser (\SI{2}{\milli\second}, \SI{100}{\nano\watt}). Grey indicates microwaves (MW) at (\SI{70}{\mega\hertz}) to counteract spin pumping. \textbf{b)} Percentage of $\vtwo$ centers whose fitted peak amplitude exceeds the count-rate threshold, shown separately for nanopillars exposed to a UV pulse (\SI{0.18}{\micro\joule} and \SI{0.24}{\micro\joule}) and those left unexposed; in each condition, $n=64$ nanopillars are examined. We find a maximum count rate for the non-exposed $\vtwo$ center of \SI{0.76}{\kilo\hertz}. For this threshold, we also observe three and eleven times more $\vtwo$ centers with the same or higher count rate at UV laser powers of \SI{0.18}{\micro\joule} and \SI{0.24}{\micro\joule}, respectively. \textbf{c}) Ensemble inhomogeneous distribution of all $\vtwo$ centers in unexposed nanopillars (bottom grey bars). A Gaussian fit yields a FWHM of the inhomogeneous distribution of \SI{22\pm1}{\giga\hertz}. The blue bars indicate the frequencies of the $\vtwo$ centers in the UV-exposed nanopillars that pass the threshold of \SI{0.3}{\kilo\hertz} (this threshold sets a cut-off to suppress bulk-$\vtwo$ contributions, also indicated by the dotted line in b)). The black dashed line shows the (scaled) Gaussian fit of the unexposed inhomogeneous distribution as a guide to the eye. \textbf{d)} Number of nanopillars that contain either 0, 1, 2, or 3 $\vtwo$ centers that pass the \SI{0.3}{\kilo\hertz} threshold. Nanopillars exposed to the UV laser are shown in shades of blue, while natural (unexposed) $\vtwo$ centers are shown in shades of grey. The numbers inside the bars show how many pillars contain that specific number of $\vtwo$ centers.} 
\end{figure*}
\subsection*{V2 concentration}
\noindent To determine whether the observed increase in PL arises from the presence of $\vtwo$ centers, we search for the characteristic zero-phonon line emission, corresponding to the A1 and A2 optical transitions, through Photoluminescence Excitation spectroscopy (PLE) spectra at \SI{4}{\kelvin} \cite{liuSiliconVacancyCenters2024,nagyHighfidelitySpinOptical2019}. The number of naturally occurring $\vtwo$ centers is then compared with the number after a single UV pulse of \SI{0.18}{\micro\joule} or \SI{0.24}{\micro\joule}. We fit all PLE spectra (see Methods and \ref{subsec:app_location_defects}) and compare the frequencies and amplitudes of the fitted peaks (example of PLE in \subfigref{fig:amound_defects}{a}). Evaluating the percentage of $\vtwo$ centers with peak amplitudes above a certain threshold reveals a significantly larger number of high-count-rate $\vtwo$ centers in UV-exposed nanopillars (see \subfigref{fig:amound_defects}{b}). Relative to the brightest $\vtwo$ center in the non-exposed nanopillars (\SI{0.76\pm0.02}{\kilo\hertz}), nanopillars exposed to a \SI{0.18}{\micro\joule} or \SI{0.24}{\micro\joule} UV pulse contain roughly three and eleven times more $\vtwo$ centers that reach or exceed this brightness, respectively. Furthermore, the brightest $\vtwo$ center found in the UV-exposed nanopillars has a count rate of \SI{2.11\pm0.04}{\kilo\hertz}, approximately a factor of 2.8 higher than the brightest emitter in the non-exposed nanopillars.

Because weak background emission can originate from defects outside the nanopillars, we restrict our analysis to $\vtwo$ centers with a PLE amplitude above \SI{0.3}{\kilo\hertz} (dotted line, \subfigref{fig:amound_defects}{a,b}). This threshold ensures that only bright, well-collected $\vtwo$ centers (most likely located near the nanopillar center, see \ref{subsec:app_collection_efficiency}) are included. 

\begin{figure*}[ht]
  \includegraphics[width=1 \textwidth]{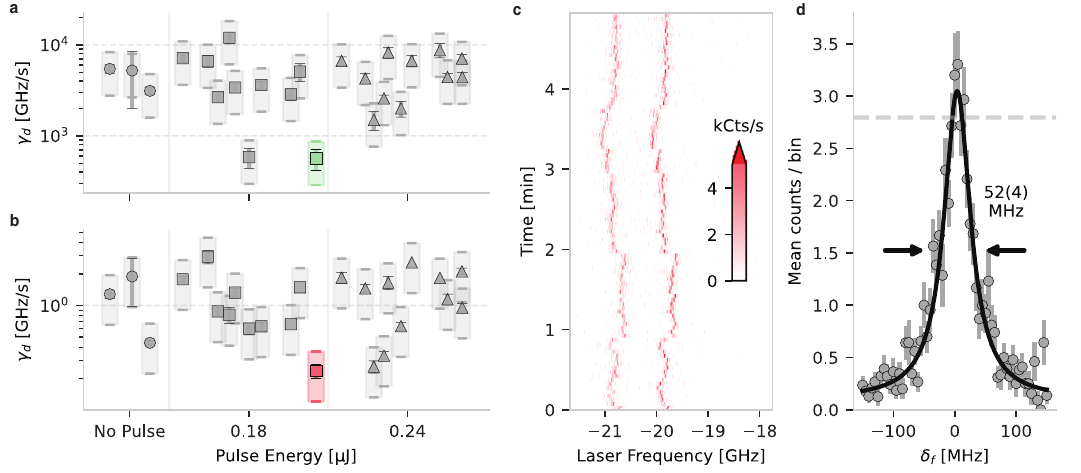}
  \caption{\label{fig:diffusion_ple} \textbf{$\vtwo$ optical properties and spectral diffusion.} \textbf{a)} Fitted spectral diffusion constant under off-resonant excitation for natural $\vtwo$ centers (circles) and laser-induced $\vtwo$ centers (squares and triangles, \SI{0.18}{\micro\joule} and \SI{0.24}{\micro\joule} respectively). We extract $\gdif$ using the methodology described in van de Stolpe \emph{et al.} \cite{vandestolpeCheckprobeSpectroscopyLifetimelimited2025}, assuming a homogeneous linewidth (FWHM) of \SI{36}{\mega\hertz}. Because the fitted $\gdif$ depends on this homogeneous linewidth (see \ref{subsec:supp_spectral_diffusion}), and this linewidth was not measured for each individual $\vtwo$ center, we indicate the spread in $\gdif$ for a plausible range of linewidths ($1\leq\Gamma/\Gamma_{\mathrm{lifetime}}\leq3$, see \ref{subsec:app_cp_ple_sweep}). $\gdif$ of the red square indicates the same $\vtwo$ center as the green square in b) and data in c) and d). \textbf{b)} Fitted spectral diffusion constants under resonant excitation. \textbf{c)} Scanning laser PLE over 5 minutes of the $\vtwo$ center highlighted in a) and b) (Pillar 14, \autoref{fig:supp_ple_location_018}). The laser is on for a total of \SI{334}{\milli\second} per scan with a power at the objective of \SI{20}{\nano\watt}. \textbf{d)} Check-Probe PLE (see van de Stolpe \emph{et al.}\cite{vandestolpeCheckprobeSpectroscopyLifetimelimited2025} for methodology) with FWHM = $\SI{52\pm4}{\mega\hertz}$ indicating similar optical linewidth ($\Gamma /\Gamma_{\mathrm{lifetime}} = \SI{2.21\pm0.04}{}$, see \autoref{fig:supp_cp_ple_sweep_018}) as natural $\vtwo$ centers in this material (see \autoref{fig:supp_cp_ple_sweep_0}). Check-probe PLE was performed at a magnetic field of \raisebox{0.4ex}{\tiny$\sim$}\SI{20}{\gauss} aligned along the crystal c-axis.}
\end{figure*}

Using the \SI{0.3}{\kilo\hertz} threshold, we find that the frequencies of the $\vtwo$ centers in the UV-exposed nanopillars follow the same ensemble inhomogeneous distribution as the natural $\vtwo$ center population (\subfigref{fig:amound_defects}{c}). This suggests that the UV pulse does not introduce significant strain or amorphisation in the vicinity of the laser-induced $\vtwo$ centers  \cite{korberFluorescenceEnhancementSingle2024,heilerSpectralStabilityV22024,udvarhelyiVibronicStatesTheir2020}. 
Furthermore, we quantify the probability of inducing only one bright $\vtwo$ center for the two UV pulse energies (\subfigref{fig:amound_defects}{d}). After a single \SI{0.18}{\micro\joule} pulse, \raisebox{0.4ex}{\tiny$\sim$}23\% of the nanopillars contain one bright $\vtwo$ center, increasing to \raisebox{0.4ex}{\tiny$\sim$}33\% for \SI{0.24}{\micro\joule}. For comparison, this is observed in only \raisebox{0.4ex}{\tiny$\sim$}6\% of non‑exposed nanopillars.
These results show that the UV-pulse strongly increases the number of $\vtwo$ centers, even though the exact origin of each $\vtwo$ center, whether created by the UV-pulse or naturally present, cannot be determined unambiguously. As we will show below, the optical properties of $\vtwo$ centers with and without UV illumination are very similar, so this ambiguity does not affect the conclusion that the UV pulse induces high-quality $\vtwo$ centers. In the remainder of this work, we will therefore refer to $\vtwo$ centers in UV-exposed nanopillars as laser-induced $\vtwo$ centers and in non-exposed nanopillars as natural $\vtwo$ centers.


\subsection*{V2 optical properties} 

\noindent Next, we characterise the optical properties of the laser-induced $\vtwo$ centers and compare them to natural $\vtwo$ centers using the check-probe spectroscopy methodology of van de Stolpe \emph{et al.} \cite{vandestolpeCheckprobeSpectroscopyLifetimelimited2025}. We characterised 24 individual $\vtwo$ centers, including 3 natural (circles in \subfigref{fig:diffusion_ple}{a,b}) and 21 laser-induced centers. Among the laser-induced centers, 10 were exposed to a single UV pulse with an energy of \SI{0.18}{\micro\joule} (squares), while 11 were exposed to a single pulse with an energy of \SI{0.24}{\micro\joule} (triangles). 

The measured spectral diffusion rates under resonant and off-resonant excitation are shown in \subfigref{fig:diffusion_ple}{a,b}. This reveals a 3-4 orders-of-magnitude difference in spectral diffusion between resonant and off-resonant laser light, consistent with previous work \cite{vandestolpeCheckprobeSpectroscopyLifetimelimited2025}. Additionally, we observe a significant spread in spectral diffusion rates, indicating that each $\vtwo$ center experiences a distinct local charge environment. Importantly, we find no significant difference in the spectral diffusion rates between laser-induced and natural $\vtwo$ centers; the two groups exhibit similar overall distributions. This suggests that the single UV pulse does not substantially alter the effective local charge environment, and that the impurities of the HPSI material or the embedding in a nanopillar, are still the main cause of spectral diffusion \cite{udvarhelyiVibronicStatesTheir2020}. Within the distribution of spectral diffusion constants ($\gdif$) for laser-induced $\vtwo$ centers, some $\vtwo$ centers remain particularly stable, exhibiting low $\gdif$ even under off-resonant excitation (green and red markers in \subfigref{fig:diffusion_ple}{a,b}).

\begin{figure*}[ht]
  \includegraphics[width=1 \textwidth]{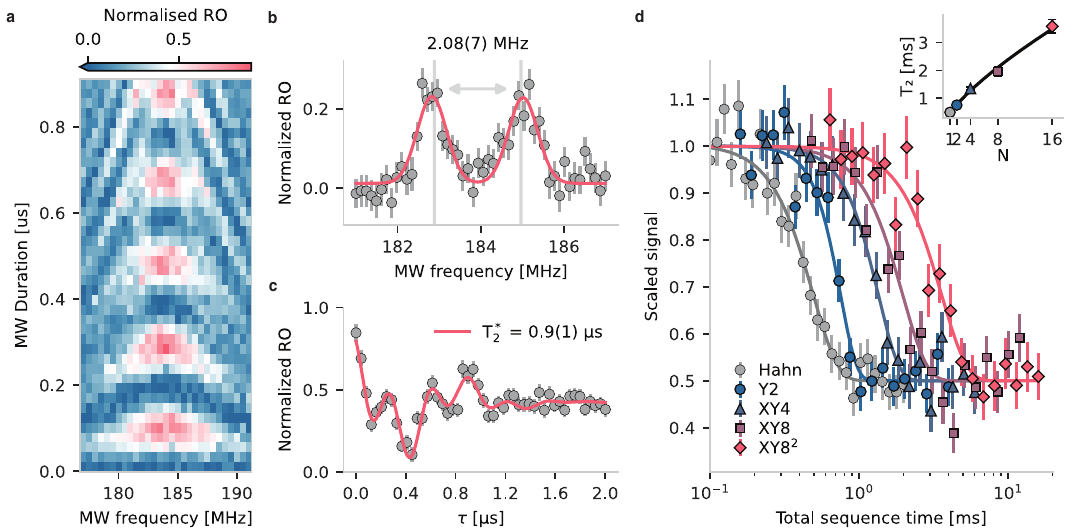}
  \caption{\label{fig:electron_spin} \textbf{Electron spin properties.} \textbf{a)} Rabi chevron pattern from the $\vtwo$ center in pillar 9. Details on the measurement sequence and normalisation procedure are elaborated in \ref{subsec:app_spin_stuff}. \textbf{b)} Electron-spin-resonance of the same $\vtwo$ center revealing a strongly coupled nuclear spin with hyperfine splitting of \SI{2.19\pm0.05}{\mega\hertz} (red line). The two solid vertical lines indicate the fitted frequencies from the Ramsey measurement, matching the ESR frequencies. \textbf{c)} Detuned Ramsey measurement on the same $\vtwo$ center with $f_{MW} = \SI{181.8}{\mega\hertz}$. Red line indicates a fit to an oscillation with two frequency components and a Gaussian decay, which translates to a hyperfine splitting of $f_{HF} = \SI{2.08\pm0.07}{\mega\hertz}$, and \Ttwostar$\,=\,$\SI{0.9\pm0.1}{\micro\second}. \textbf{d)} Scaled Hahn-echo ($N=1$) and dynamical decoupling ($N =2 $ to $N =16$) measurements on the $\vtwo$ center in pillar 14 (\autoref{fig:supp_ple_location_018}) with no (resolvable) strongly coupled nuclear spin (see \autoref{fig:supp_spin_4_32} for Chevron, ESR and Ramsey) at a magnetic field of \raisebox{0.4ex}{\tiny$\sim$}\SI{1300}{\gauss}. We find a \Ttwohahn = \SI{0.49\pm0.02}{\milli\second} and extend the coherence time to \Ttwodd = \SI{3.6\pm0.3}{\milli\second} with two $\mathrm{XY8}$ sequences. The inset shows $\Ttwo$ as a function of the number of $\pi$-pulses N, fitted to a power function $\Ttwo=\mathrm{\beta\cdot N^{\alpha}}$(excluding Hahn-echo) from which we extract $\alpha = \SI{0.73\pm0.04}{}$ and $\beta = \SI{0.46\pm0.04}{\milli\second}$.}
\end{figure*}

We perform two additional types of photoluminescence excitation (PLE) measurements. First, we observe the spectral wandering (\subfigref{fig:diffusion_ple}{c}) by scanning a resonant laser continuously whilst applying microwaves; second, we use a check-probe type PLE on the A1 transition to approximate the homogeneous linewidth (\subfigref{fig:diffusion_ple}{d}) \cite{vandestolpeCheckprobeSpectroscopyLifetimelimited2025}. 
By sweeping the threshold for the check-probe PLE, we can minimize the residual broadening due to imperfect initialisation and compare this homogeneous linewidth to the lifetime-limited linewidth (see \ref{subsec:app_cp_ple_sweep}). We find linewidths satisfying $1\leq\Gamma/\Gamma_{lifetime}\leq4$ for all measured $\vtwo$ centers, indicating that there is still some broadening (either residual (in)homogeneous broadening, or power broadening). The linewidths for the natural $\vtwo$ centers are similar to the laser-induced $\vtwo$ centers, indicating that the UV pulse does not significantly lower the optical coherence of the $\vtwo$ centers and, importantly, are compatible with remote entanglement generation under modest time filtering \cite{legeroTimeresolvedTwophotonQuantum2003}.

\subsection*{V2 electron-spin properties} 

\noindent In addition to optical coherence, the electron spin coherence is a crucial component for utilising laser-induced $\vtwo$ centers in sensing, quantum information processing, and quantum network applications. We investigate the spin properties of two single laser-induced $\vtwo$ centers, which exhibits relatively low spectral diffusion ($\gdif$) under both resonant and off-resonant excitation (pillar 9 and 14; see \autoref{fig:supp_ple_location_018} for location details, \autoref{fig:app_diffusion_spin_v2} for optical properties of $\vtwo$ center in pillar 9, \subfigref{fig:diffusion_ple}{c,d} for optical properties of $\vtwo$ center in pillar 14 and \autoref{fig:supp_spin_4_32} for spin properties of $\vtwo$ center in pillar 14). Note that this optical selection might already introduce a bias in the electron spin environment and, thus, potentially in the electron spin coherence.

First, we operate at a magnetic field of \raisebox{0.4ex}{\tiny$\sim$}\SI{40}{\gauss}. This ensures that the Kramers degeneracy within the $\pm\frac{1}{2}$ and $\pm\frac{3}{2}$ spin eigenstates is lifted, and that the ground or excited state level crossings are sufficiently detuned\cite{soykalSiliconVacancyCenter2016,siminAllOpticalDcNanotesla2016,udvarhelyiVibronicStatesTheir2020}. \subfigref{fig:electron_spin}{a} shows Rabi oscillations for driving the ground state $\ket{m_s=+\frac{1}{2}} \leftrightarrow \ket{m_s=+\frac{3}{2}}$ transition of the $\vtwo$ center found in pillar 9. On the same $\vtwo$ center, we perform electron spin resonance (ESR), see \subfigref{fig:electron_spin}{b} (details of the sequence can be found in \ref{subsec:app_spin_stuff}). ESR reveals a strongly coupled nuclear spin, with a hyperfine coupling of \SI{2.19\pm0.05}{\mega\hertz}. This coupling is consistent with either a coupled $^{13}$C or $^{29}$Si nuclear spin as previously observed and reported \cite{hesselmeierQuditBasedSpectroscopyMeasurement2024}, and accounts for the detuning visible in the Rabi chevron pattern (\subfigref{fig:electron_spin}{a} and \ref{subsec:app_spin_stuff}). We perform Ramsey measurements to extract the spin-dephasing time \Ttwostar, finding \Ttwostar=\SI{0.9\pm0.1}{\micro\second} (\subfigref{fig:electron_spin}{c}).

Next, we operate at a higher magnetic field (\raisebox{0.4ex}{\tiny$\sim$}\SI{1300}{\gauss}) to suppress decoherence from anisotropic hyperfine interactions with the spin bath. We perform Hahn-echo and dynamical decoupling (DD) measurements on the $\vtwo$ center in pillar 14, which has no strongly coupled nuclear spins. We measure a Hahn-echo coherence time of $\Ttwohahn = \SI{0.49\pm0.02}{\milli\second}$ and extend the coherence up to $\Ttwodd = \SI{3.6\pm0.3}{\milli\second}$ using two consecutive $\mathrm{XY8}$ sequences (\subfigref{fig:electron_spin}{d}).

The $\Ttwodd$ values display a power-law dependence on the number of decoupling pulses, with a fitted exponent of $\alpha = \SI{0.73\pm0.04}{}$. This is slightly higher than the $\alpha=2/3$ scaling associated with a Lorentzian noise spectrum ($\mathrm{\raisebox{0.4ex}\tiny\sim\frac{1}{\omega^2}}$), characteristic of Ornstein-Uhlenbeck dynamics \cite{delangeUniversalDynamicalDecoupling2010}, indicating a slight deviation from a single-Lorentzian noise spectrum toward lower frequency noise\cite{medfordScalingDynamicalDecoupling2012}. 
Despite the relatively high nitrogen concentration of  (\raisebox{0.4ex}{\tiny$\sim$}\SI{1.1e15}{\per\centi\meter\cubed}, see Methods) our $\Ttwostar$ and $\Ttwohahn$ are comparable to values reported for natural-isotope-abundance epitaxially grown 4H-SiC with doping levels two orders of magnitude lower\cite{laiSingleShotReadoutNuclear2024,nagyQuantumPropertiesDichroic2018,siminLockingElectronSpin2017,yangElectronSpinDecoherence2014,kunaLocalizationCoherentControl2025,steidlSingleV2Defect2025}. This suggests that neither the doping level nor the laser-writing method limits the spin coherence. These results show that commercially available wafers combined with laser-induced defect creation might provide a viable path towards quantum technologies based on colour centers in 4H-SiC.

\section*{Discussion}
\noindent In this work, we studied the creation of laser-induced $\vtwo$ centers in 4H-SiC nanopillars in bulk-grown SiC. We observe narrow optical linewidths that are compatible with remote entanglement generation\cite{legeroTimeresolvedTwophotonQuantum2003}, and long electron-spin coherence times consistent with dephasing by the nuclear spin bath\cite{steidlSingleV2Defect2025}, although potential contributions from electron spin impurities cannot be excluded.

Looking forward, while here we focus on the formation of V2 centers in bulk-grown 4H-SiC wafers, the methods employed here can potentially be extended to high-quality, isotopically engineered, epitaxially grown 4H-SiC \cite{bourassaEntanglementControlSingle2020,bulancea-lindvallIsotopePurificationInducedReductionSpinRelaxation2023,parthasarathyScalableQuantumMemory2023,lekaviciusOrdersMagnitudeImprovement2022,marcksNuclearSpinEngineering2025}, and as well to other defect centers, such as V1 centers (h-site $\vsi$) \cite{nagyHighfidelitySpinOptical2019}, divacancies \cite{andersonFivesecondCoherenceSingle2022}, and non-native atom-related complexes like NV centers \cite{normanSub2KelvinCharacterization2025}. A systematic exploration of defect formation, density, and properties using above‑bandgap laser writing, across different irradiation conditions and writing parameters, can enable deeper insight into the underlying material science and the mechanisms of defect creation, including potential local annealing or activation effects \cite{kuatedefoEnergeticsKineticsVacancy2018}, in different types of SiC and nanostructures.

Additionally, monitoring the photoluminescence between laser pulses might enable the deterministic creation of a desired number of defects \cite{chenLaserWritingIndividual2019}. Large-scale fabrication methods, such as Silicon-Carbide-on-Insulator \cite{lukin4HsiliconcarbideoninsulatorIntegratedQuantum2020} or angled etching \cite{majetyWaferscaleIntegrationFreestanding2025}, would open the door to complex integrated nanophotonic structures, including photonic crystal cavities. 
Together with the laser-induced creation of high-quality $\vtwo$-centers in commercial bulk-grown SiC wafers presented here, this combination might enable a path towards efficient, high-quality defect creation in scalable quantum devices. 

\section*{Methods}

\subsection*{Sample preparation}
\noindent The sample was diced directly from a 6-inch High-Purity Semi-Insulating (HPSI) wafer obtained from Wolfspeed (model type W4TRG0R-N-0200). We note that the HPSI terminology originates from the silicon carbide electronics industry. In the quantum technology context considered here, this material has a significant amount of residual impurities (order \raisebox{0.4ex}{\tiny$\sim$}$10^{15}\,\SI{}{\per\centi\meter\cubed}$ according to Son et al.\cite{sonChargeStateControl2021}) and is hence considered low material quality with respect to a concentration of \raisebox{0.4ex}{\tiny$\sim$}$10^{13}\,\SI{}{\per\centi\meter\cubed}$ typical for epitaxially grown layers on the c-axis of silicon carbide \cite{nagyHighfidelitySpinOptical2019,heilerSpectralStabilityV22024}. On a different sample, diced from a wafer of the same model type, secondary-ion mass spectrometry (SIMS) determined the nitrogen donor concentration as [N] = $\SI{1.1e15}{\per\centi\meter\cubed}$. In addition to intrinsic silicon vacancies, additional silicon vacancies were generated in this sample through a previous (unrelated) 2 MeV electron irradiation with a fluence of $\SI{2e13}{\per\centi\meter\squared}$. The sample was annealed at \SI{600}{\celsius} for 30 min in an Argon atmosphere. To enhance the optical collection efficiency and mitigate the unfavourable $\vtwo$ dipole orientation for confocal access along the SiC growth axis (c-axis), we utilize nanopillars. Nanopillar fabrication begins with sputtering an aluminium oxide (Al$_{2}$O$_{3}$) mask on the SiC substrate, followed by spinning and patterning of AR-P 6200 e-beam resist (Allresist GmbH). The resist pattern is transferred into the Al$_{2}$O$_{3}$ using inductively coupled plasma reactive ion etching (ICP-RIE) with a BCl$_{3}$/Cl$_{2}$ plasma. Afterwards, the pattern is etched into the SiC using ICP-RIE and an SF$_{6}$/O$_{2}$ plasma. After ICP-RIE, the remaining Al$_{2}$O$_{3}$ was removed with hydrofluoric acid.

\subsection*{Pulsed UV experiment}
\noindent All pulsed UV experiments are performed at room temperature and ambient pressure.
To determine the LIAT for bulk material, we define a grid of 100 spots (10 by 10) for each pulse energy. To avoid spatial and temporal biasing (e.g., alignment or temperature drifts), the pulse energies are varied in a randomized order and calibrated before each measurement. All spots are positioned within a region where alignment artifacts are negligible. The z-position of the objective is fine-tuned by scanning through different heights while applying a pulse energy slightly above the LIAT. We then use the z position that produces the largest amorphization on the surface (see \ref{subsec:app_damage_threshold}).

The same procedure is used to determine the LIAT for nanopillars, with the grid consisting of 4 by 25 individual structures. The measurement process is fully automated, involving piezo movement of the objective and a single UV pulse. Due to grid interpolation and piezo positioner hysteresis, the UV pulse may not always focus in the exact center of each nanopillar. While this misalignment has minimal effect on the LIAT determination, it can slightly influence the resulting concentration of $\vtwo$ centers.

\subsection*{Automated PLE V2 selection}
\noindent We measure PLE spectra on 192 nanopillars and fit them automatically. We apply microwaves to the $\vtwo$ centers using a bondwire (\raisebox{0.4ex}{\tiny$\sim$}\SIrange{20}{50}{\micro\meter} away from the laser-induced $\vtwo$ centers) to efficiently counter spin-pumping.

For the non-UV-exposed and UV-exposed nanopillars, we perform the following measurement sequence.
\begin{enumerate}
    \item Optimize the x,y,z position of the focal point to maximize the amount of counts under off-resonant excitation for each nanopillar.
    \item Perform photoluminescence excitation (PLE) measurement by scanning a resonant laser over 60 GHz (while continuously applying microwaves at 70 MHz) centered around the observed inhomogeneous distribution of  $\vtwo$ centers.
    \item Fit a Voigt profile to all observed peaks in the PLE spectra and store the center frequency and amplitude of the fit. To avoid double counting (A1 and A2 transition of a single $\vtwo$ center) and fitting errors, we set the additional constrains to the fit: distance between fitted peaks >\SI{2}{\giga\hertz} and the FWHM of the fit should be larger than \SI{50}{\mega\hertz} and lower than \SI{3000}{\mega\hertz} and have a minimal count rate of \SI{0.03}{\kilo\hertz}.
\end{enumerate}

\subsection*{Experimental setup}
\noindent All room temperature experiments are performed using a home-built confocal microscopy setup. We use a \SI{3}{\nano\second} pulsed UV (\SI{337}{\nano\meter}) laser (Lasertechnik Berlin, MNL103-LD) together with a \SI{0.7}{NA} objective (LUCPLFLN60X) for defect formation experiments. For PL measurements, the objective is switched to a \SI{0.9}{NA} objective (Olympus MPLFLN100X).  The UV excitation path is separated from the V2 excitation and collection pathway using a dichroic mirror (Semrock Di01-R355). V2 centers are repumped with a 785 nm laser (Cobolt 06-MLD785), and the resulting photoluminescence is filtered through an 830 nm long-pass filter (Semrock BLP01-830R-25) before being detected by superconducting nanowire single-photon detectors (SNSPDs, Single Quantum).

Low-temperature experiments are performed using a home-built confocal microscopy setup at 4K (Montana Instruments S100). The NIR lasers (Toptica DL Pro and the Spectra-Physics Velocity TLB-6718-P) are frequency-locked to a wavemeter (HF-Angstrom WS/U-10U) and their power is modulated by acousto-optic modulators (G\&H SF05958). A wavelength division multiplexer (OZ Optics) combines the \SI{785}{\nano\meter} repump (Cobolt 06-MLD785) and NIR laser light, after which it is focused onto the sample using a fast-steering mirror (Newport FSM-300) and a movable, room temperature objective (Olympus MPLFLN 100X), which is kept at a vacuum and is thermally isolated by a heat shield. A 90:10 beam splitter that directs the laser light into the objective, allows $\vtwo$ center phonon-sideband emission to pass through, to be detected on a superconducting nanowire single photon detector (SNSPD, Single Quantum, filtered by a Semrock FF01-937/LP-25 long pass filter at a slight angle). Alternatively, the emission can be directed to a spectrometer (Princeton Instruments IsoPlane 81) and filtered by an \SI{830}{\nano\meter} long-pass filter (Semrock BLP01-830R-25). Microwave pulses are generated by an arbitrary-waveform generator (Zurich Instruments HDAWG8) and an RF source (R\&S SGS100A SGMA RF source), amplified (Mini-circuits LZY-22+ or ZHL-15W-422-S+), and delivered with a bond-wire drawn across the sample. The coarse-time scheduling ($\SI{1}{\micro \second}$ resolution) of the experiments is managed by a microcontroller (ADwin Pro II). For a schematic of the room-temperature and low-temperature setups, see \ref{subsec:app_optical_setup}. All data was taken using the QMI Python package \cite{teraaQMIQuantumMeasurement2026}.

\subsection*{Magnetic field}

\noindent For the check-probe PLE in \subfigref{fig:diffusion_ple}{d} and spin coherence measurements  \subfigref{fig:electron_spin}{a,b,c}, we apply a low magnetic field along the defect symmetry axis (c-axis) using a permanent neodymium magnet outside the cryostat. For the check-probe PLE, we used a magnetic field of approximately \raisebox{0.4ex}{\tiny$\sim$}\SI{20}{\gauss}, and for the spin coherence measurements, we applied a magnetic field of approximately \raisebox{0.4ex}{\tiny$\sim$}\SI{40}{\gauss} so that the $\ket{m_s=+\frac{1}{2}} \leftrightarrow \ket{m_s=+\frac{3}{2}}$ transition is around \SI{181.8}{\mega\hertz}. The magnetic field was aligned by applying the sequence depicted in the inset of \subfigref{fig:amound_defects}{a} without applying MW pulses. We set $f_1$ to be resonant with the A2 transition and monitor the average photoluminescence ($f_1$ pulse duration is \SI{2}{\milli \second}). A (slightly) misaligned field causes spin-mixing between the $m_s = \pm \tfrac{3}{2}$ and $m_s=\pm \tfrac{1}{2}$ subspace, which increases the detected signal. Minimising the photoluminescence thus optimises the field alignment along the symmetry axis. 

For the \Ttwodd in \subfigref{fig:electron_spin}{d} we apply a magnetic field of \raisebox{0.4ex}{\tiny$\sim$}\SI{1300}{\gauss} using a permanent neodymium magnet just behind the sample. We place 3 external magnets outside the cryostat on motorized stages to align the magnetic field as described previously. 

\section*{Data availability}
\noindent All data underlying the study will be made available on the open 4TU data server.

\section*{Code availability}
\noindent Code used to operate the experiments is available on request.

\section*{Acknowledgements}
\noindent We thank Jonathan Dietz for the useful discussions on above-bandgap laser writing. We thank Vadim Vorobyov and Florian Kaiser for the input on the optical setup and general discussions. We thank Benjamin van Ommen, Jiwon Yun, Margriet van Riggelen and Nicolas Demetriou for discussions on spin dynamics and control. We thank Hans Beukers and Christopher Waas for their help with automating measurements.
We thank Nico Albert and Tim Hiep for designing and machining parts for the setup, Henri Ervasti, Pieter Botma and Ravi Budhrani for software support and development, Régis Méjard and Hitham Mahmoud Amin for optical and safety support, Siebe Visser, Vinod Narain and Erik van der Wiel for general technical support and Jason Mensigh, Olaf Benningshof and Jared Croese for cryogenic and vacuum engineering support.

This project has received funding from the European Research Council (ERC) under the European Union’s Horizon 2020 research and innovation programme (grant agreement No. 852410). 
This work was supported by the Dutch National Growth Fund (NGF), as part of the Quantum Delta NL programme. 
This work is part of the research programme NWA-ORC with project number NWA.1160.18.208, which is (partly) financed by the Dutch Research Council (NWO).
This project has received funding from the European Union’s Horizon Europe research and innovation program under grant agreement No
 101135699. 
 This work was funded by the European Union's Horizon Europe research and innovation programme under grant agreement No. 101102140 – QIA Phase 1. 
 This project was funded within the QuantERA II Programme that has received funding from the EU H2020 research and innovation programme under GA No 101017733
 
\section*{Author contributions}
\noindent LJF, GMT and RK performed the room temperature creation and detection experiments with analytical and discussion support from AMD and ELH. GMT and RK performed the amorphisation threshold measurements.  GMT designed and fabricated the sample. LJF, YH, and TM performed the low-temperature optical and spin property measurements. GvdS, SJH, LJF, GMT, RK, TD, and AD built the experimental apparatus. LJF, GMT, RK, YH and THT analysed the data. LJF, GMT, YH and THT wrote the manuscript with input from all authors. THT supervised the project.

\section*{Competing interests}
\noindent The authors declare no competing interests.

\bibliography{SiC_Laser_Writing_last}

\clearpage
\newpage
\appendix
\renewcommand{\thesubsection}{Supplementary Note~\arabic{subsection}}
\renewcommand{\subsectionautorefname}{Supplementary Note}
\renewcommand{\sectionautorefname}{Supplementary Note}

\clearpage
\widetext
\begin{center}
\textbf{\large Supplementary Information for \\Laser-induced creation of coherent V2 centers in bulk-grown silicon carbide}
\end{center} 

\renewcommand{\figurename}{\textbf{Supplementary Figure}}
\renewcommand{\tablename}{\textbf{Supplementary Table}}
\renewcommand{\thesubsection}{Supplementary Note \arabic{subsection}}
\renewcommand{\subsectionautorefname}{Supplementary Note}
\renewcommand{\sectionautorefname}{Supplementary Note}
\renewcommand{\thefigure}{S\arabic{figure}}
\renewcommand{\figurename}{Figure} 
\setcounter{subsection}{0}

\setcounter{equation}{0}
\setcounter{figure}{0}
\setcounter{table}{0}
\setcounter{page}{1}
\makeatletter
\renewcommand{\theequation}{S\arabic{equation}}


\subsection{Laser-induced amorphisation threshold}
\label{subsec:app_damage_threshold}

\begin{figure}[ht]
   \centering
  \includegraphics[width=1 \textwidth]{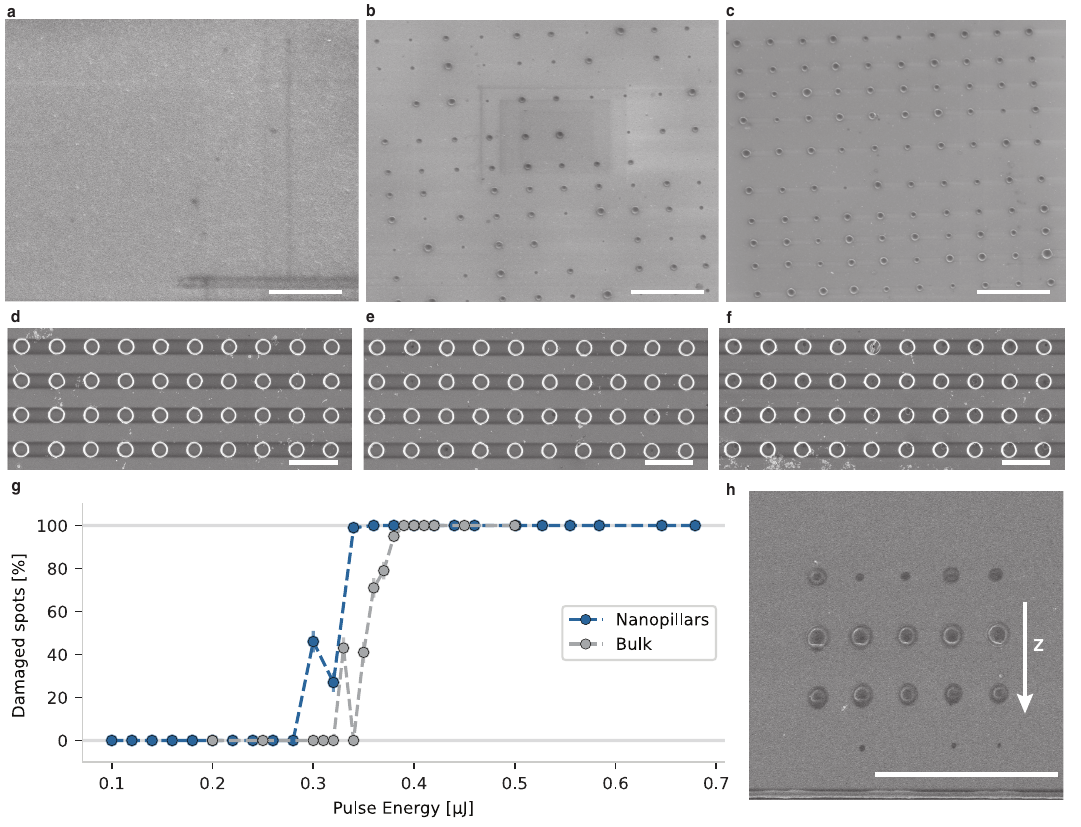}
  \caption{\textbf{SEM of LIAT for bulk and nanopillars.} \textbf{a)} Bulk experiment with single pulse energy of \SI{0.20}{\micro\joule}, of a grid of 100 spots with no visible amorphisation. \textbf{b)} Bulk experiment with single pulse energy of \SI{0.38}{\micro\joule} with probabilistic (80\%) amorphisation. \textbf{c)} Bulk experiment with single pulse energy of \SI{0.40}{\micro\joule} with 100\% amorphisation. \textbf{d)} Nanopillar experiment with single pulse energy of \SI{0.24}{\micro\joule} of a grid of 100 spots with no visible amorphisation. \textbf{e)} Nanopillar experiment with single pulse energy of \SI{0.3}{\micro\joule} with probabilistic (46\%) amorphisation. \textbf{f)} Nanopillar experiment with single pulse energy of \SI{0.34}{\micro\joule} with 100\% amorphisation. \textbf{g)} The number of damaged sites for each pulse energy, for both bulk and nanopillars. \textbf{h)} Example of the focusing procedure of the UV laser. For each row, the z-position is slightly adjusted to compare the surface-damage diameter. In all the SEM images, the white bar indicates \SI{5}{\micro\meter}.}
  \label{fig:supp_damage_np_bulk}
\end{figure}

Due to the efficient generation of electron-hole pairs using an above-bandgap laser, linear photo-ionization can generate a high density of free electrons. This can lead to the formation of a highly ionised plasma, potentially resulting in material damage \cite{woodLaserInducedDamageOptical2003a,schafferLaserinducedBreakdownDamage2001,ballingLaserCouplingRelaxation2020}. Given the fixed pulse duration of \SI{3}{\nano\second}, we systematically vary the pulse energy of single-shot exposures to determine the damage threshold. This threshold defines the upper bound of pulse energy used in subsequent investigations, referred to as the laser-induced amorphisation threshold (LIAT). The procedure is applied to both bulk 4H-SiC and nanostructured pillars (diameter \raisebox{0.4ex}{\tiny$\sim$}\SI{1.2}{\micro\meter}, height \raisebox{0.4ex}{\tiny$\sim$}\SI{1}{\micro\meter}) and followed by SEM inspection, see \subfigref{fig:supp_damage_np_bulk}{a-f}.

Following SEM analysis, the number of damaged sites was quantified and plotted as a function of pulse energy (\subfigref{fig:supp_damage_np_bulk}{g}), revealing the energy dependence of laser-induced damage thresholds across different surface morphologies. A distinctly non-linear response is observed in both the bulk material and the nanopillars. As we do not focus on the probabilistic armophization regime, we do not investigate this further.

Prior to each pulse-energy configuration, the UV laser focus was optimised on the substrate surface by adjusting the objective's z-axis, while using high-energy UV pulses of \SI{0.45}{\micro\joule}, to maintain consistent exposure parameters. As can be seen in \subfigref{fig:supp_damage_np_bulk}{h}, the second row exhibits the largest surface-damage diameters, which we attribute to optimal focusing of the UV laser at the substrate surface during the experiment.

\subsection{Simulated collection efficiency}
\label{subsec:app_collection_efficiency}
\begin{figure}[ht]
  \includegraphics[width=1 \textwidth]{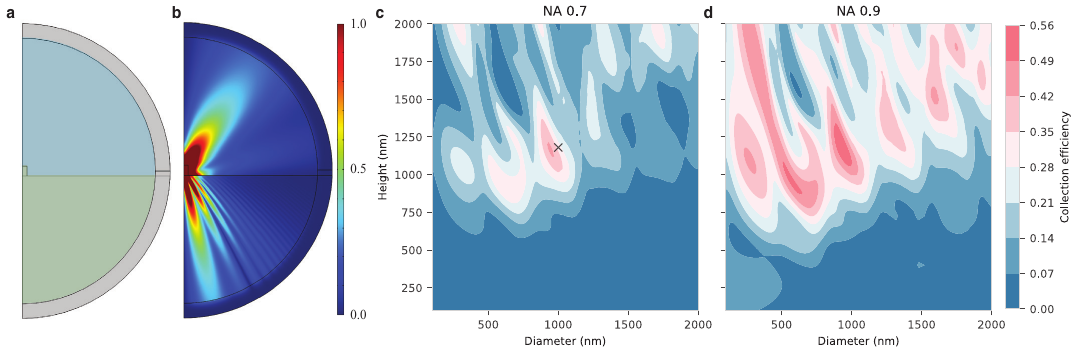}
  \caption{\textbf{Simulated collection efficiency for nanopillars.} \textbf{a)} Geometry used to simulate the V2 center ZPL emission pattern in nanopillars using COMSOL Multiphysics 5.6. 2D axis-symmetry is applied to reduce computational time. The green area is specified as 4H-SiC (\cite{singhNonlinearOpticalProperties1971}: Î±-SiC; n(o) 0.488-1.064 µm), the blue area is specified as air, and the outer grey area is defined as a perfectly matched layer. In the center of the nanopillar, we place an artificial copper antenna to emulate the V2 center ZPL emission. \textbf{b)} The emission pattern of the V2 center in a nanopillar with dimensions as indicated with the black cross in \textbf{c}. For visualization purposes, the colour scale has been saturated to enhance the contrast and reveal the underlying emission pattern. \textbf{c),d)} The calculated collection efficiency for various nanopillar diameters and heights with a 0.7 and 0.9 NA objective, respectively.} 
  \label{fig:supp_collection}
\end{figure}
The perpendicular optical dipole orientation of the V2 defect with respect to the surface (in c-plane 4H-SiC) limits the collection efficiency through an objective positioned above the sample. Nanopillars are an effective way to alleviate this limitation. To simulate the effects of a nanopillar structure, we use the 2D-axisymmetric electromagnetic waves, frequency-domain physics interface of COMSOL Multiphysics 5.6. \subfigref{fig:supp_collection}{a} illustrates the geometry used. The green region is specified as 4H-SiC (Singh et al. 1971: Î±-SiC; n(o) 0.488-1.064 µm), the blue region is air, and the outer grey area is a perfectly matched layer. In the center of the nanopillar, an artificial copper antenna models the optical emission of a V2 defect, with a single frequency emission corresponding to the ZPL transition. The emission of a V2 defect in a nanopillar (dimensions indicated with the black cross in \subfigref{fig:supp_collection}{c}) are shown in \subfigref{fig:supp_collection}{b}. 
The collection efficiencies in \subfigref{fig:supp_collection}{c,d} are calculated with:
\begin{equation}
    \eta = \frac{P_\mathrm{objective}}{P_\mathrm{tot}},
\end{equation}
with $P_{tot}$ being the total power emitted by the artificial antenna and $P_{obj}$ the power directed towards the objective, calculated as:
\begin{equation}
    P_\mathrm{obj} = \int_0^{\sin^{-1}(\text{NA})}P_\mathrm{boundary}(\theta)\mathrm{d}\theta,
\end{equation}
where $P_\mathrm{boundary}$ is the power normal to the boundary surface and $\text{NA} = 0.7$ and $\text{NA} = 0.9$ in \subfigref{fig:supp_collection}{c,d}, respectively.\\

These simulations are performed only at the ZPL wavelength and with the emitter positioned at the center of the nanopillar. While a real $\vtwo$ defect emits over a broader spectrum and may reside at different locations within the pillar, exploring these additional configurations lies beyond the scope of this work. The present model is therefore intended as a representative estimate to guide the nanopillar geometry rather than an exhaustive optimization.

\subsection{2D-PL Before and after UV pulse}
\label{subsec:app_before_after_2dpl}
\begin{figure}[ht]
  \includegraphics[width=1 \textwidth]{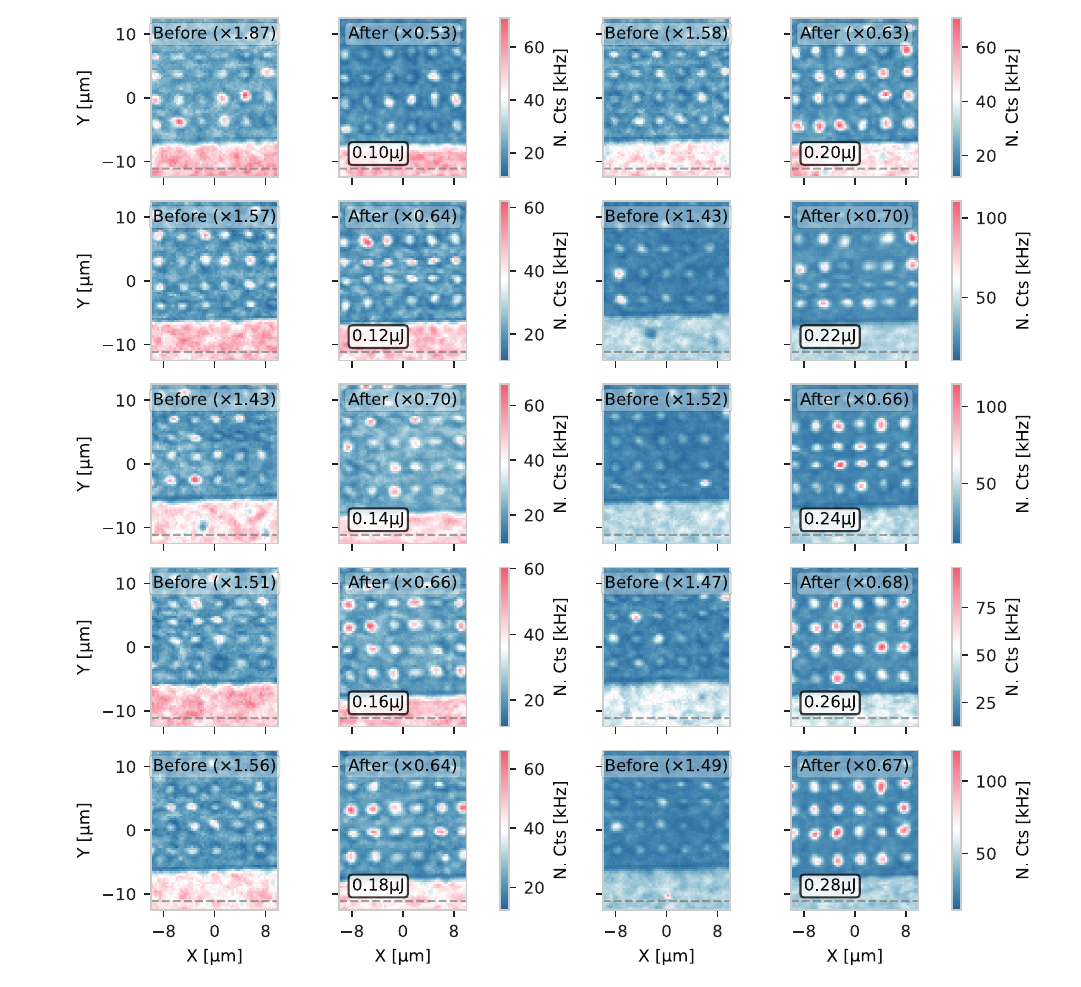}
  \caption{\textbf{Normalised 2D PL before and after a single UV pulse for different pulse energies}. Here, the factor $\beta$ is indicated in each 2D PL scan. The dashed grey line indicates the Y location, which is used to normalise the before- and after-PL counts with respect to each other.} 
  \label{fig:supp_2d_before_after}
\end{figure}

We perform 2D photoluminescence (PL) experiments on all nanopillars, followed by single UV pulse exposure, with varying pulse energy. Subsequently, we perform 2D PL scans again over the same nanopillars.\\
The 2D‑PL measurements are performed with a different objective than the one used for the UV pulse, and exchanging objectives leads to a small shift in collection efficiency. Therefore, in order to have a fair comparison between the before- and after-scans, we use the mean count rate of a reference bulk region (grey dashed line in \ref{fig:supp_2d_before_after}) to establish a common baseline and extract the appropriate scaling factor between the two measurements. We use $\alpha = \sqrt{\mu_\mathrm{before}\times \mu_\mathrm{after}}$ (with $\mu$ the mean along the dashed grey line) and define the scaling factor for before and after as $\beta_\mathrm{before,after} = \frac{\mu_\mathrm{before,after}}{\alpha}$. We multiply all the counts in the before- and after-scans by $\beta_\mathrm{before}$ and $\beta_\mathrm{after}$, respectively.
Note that more nanopillars were exposed to a UV pulse than the \raisebox{0.4ex}{\tiny$\sim$}24 visible in each 2D‑PL scan, and not all scans are included in this supplement. The scans shown are taken at approximately the same X–Y positions, although the exact center positions before and after exposure do not perfectly coincide.

At pulse energies of \SI{0.16}{\micro\joule} and above, we observe a clear increase in PL across multiple nanopillars, marking the onset of light-emitting defect formation. As the LIAT is approached, all nanopillars exhibit a substantial PL enhancement, suggesting the formation of additional emitters or non‑visible (with SEM inspection) amorphization.

\subsection{Room-temperature laser saturation}
\label{subsec:app_rt_violin}
\begin{figure}[ht]
\centering
  \includegraphics[width=0.5 \textwidth]{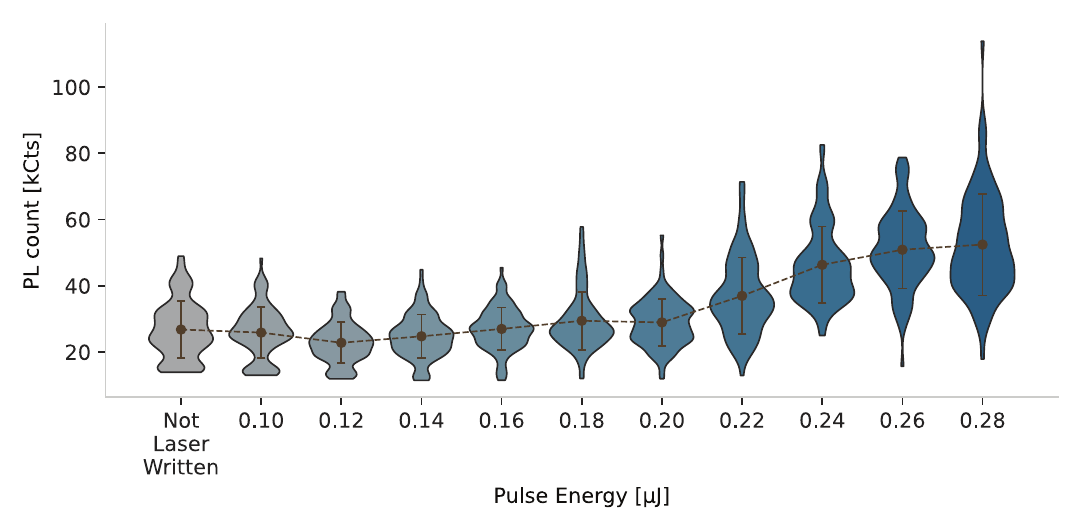}
  \caption{\textbf{Distribution Photoluminescence:} Maximum PL signal after scanning X, Y and Z axis of the objective with an off-resonant laser power of \SI{500}{\micro\joule}. The violin plot displays the distribution (symmetric histogram) for each UV pulse energy.} 
  \label{fig:supp_PL_violin}
\end{figure}

\begin{figure}[ht]
\centering

  \includegraphics[width=0.5 \textwidth]{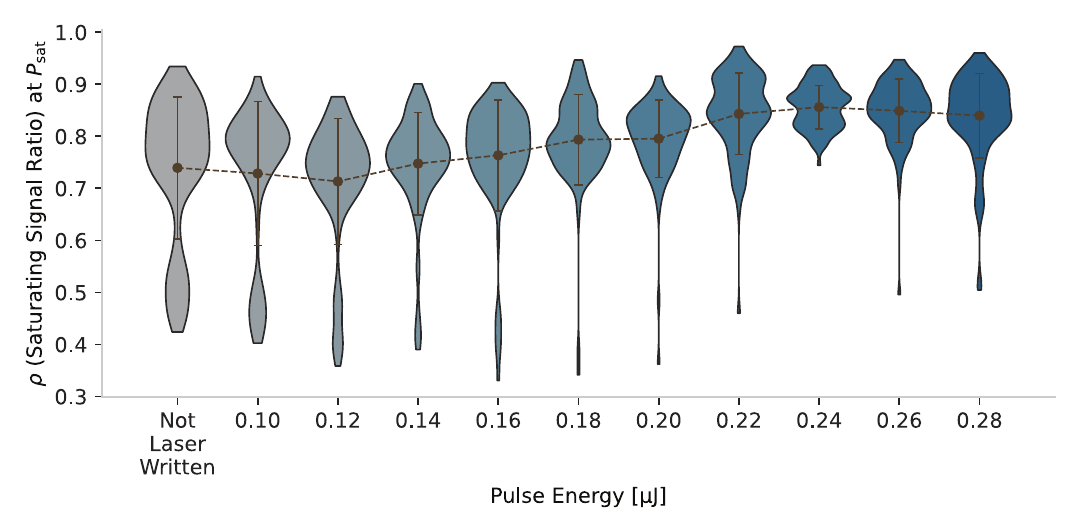}
  \caption{\textbf{Signal vs. background ($\rho$) at saturation power.} Violin plot distribution of the fraction between the amplitude of the saturating signal vs the background signal at fitted saturation power.}
  \label{fig:supp_rho_violin}
\end{figure}

At room temperature, we performed PL measurements on all UV-exposed nanopillars (100 nanopillars per pulse energy). For each nanopillar, we first optimize the objective position by applying \SI{500}{\micro\watt} of off-resonant laser power and scan all 3 axis of the objective. Afterwards, we extract the PL counts at the optimal position and plot the distribution of maximum PL counts versus pulse energy in a violin plot \autoref{fig:supp_PL_violin}. We observe that the level of PL counts and spread increases for pulse energies $\geq \SI{0.18}{\micro\joule}$.
 Next, we sweep the off-resonant laser power from \SIrange{0}{2}{\milli\watt} and collect the PL (example in main text \subfigref{fig:defect_creation}{f}). We fit the data to a saturation curve (PL from defect) + a linear term (background PL):
\begin{equation}
    PL = B\cdot p + A\cdot\frac{p}{p+p_\mathrm{sat}}.
\end{equation}
Here, $p$ indicates the off-resonant laser power and $B, A$ and $p_\mathrm{sat}$ are fitting parameters. We then define $\rho$ at $p_\mathrm{sat}$ as $\frac{A_\mathrm{sat}}{A_\mathrm{sat}+B_\mathrm{sat}}$, with $A_\mathrm{sat} = \frac{A}{2}$ and $B_\mathrm{sat} = B\cdot p_\mathrm{sat}$ which gives us the ratio between the signal and the background PL. We plot the distribution of $\rho$ at $p_{sat}$ for each UV pulse energy in the violin plots displayed in \autoref{fig:supp_rho_violin}. We observe a decrease in the lower tail of the distribution ($\rho \leq 0.6$) for an increase in UV pulse energy. Furthermore, we notice a slight increase of $\rho$ at $p_\mathrm{sat}$ and a small decrease in variance for pulse energies $\geq \SI{0.16}{\micro\joule}$. However, there is also a slight decrease in $\rho$ for pulse energies $\geq \SI{0.24}{\micro\joule}$, hinting at an increase in background fluorescence. 

\subsection{Location of V2 centers}
\label{subsec:app_location_defects}

\begin{figure}[H]
  \includegraphics[width=1 \textwidth]{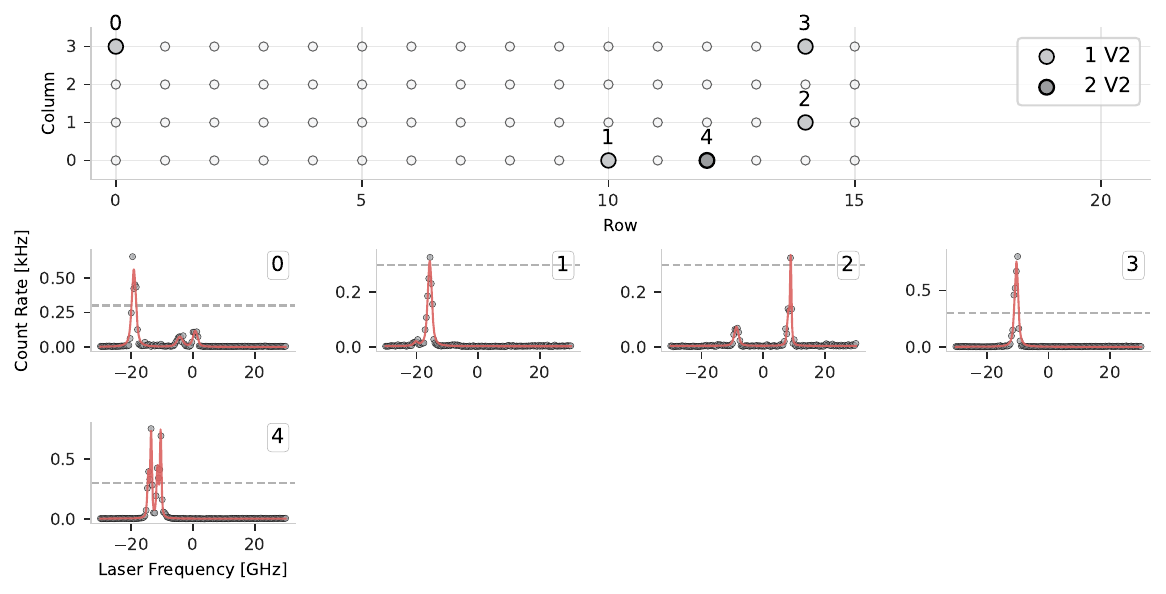}
  \caption{\textbf{Photoluminescence excitation of 64 unexposed nanopillars.} The white circles indicate a raster of nanopillars. The grey circles indicate nanopillars with V2 centers passing the \SI{0.3}{\kilo\hertz} threshold (grey dashed line), and their respective PLEs are shown below. Note that for all the PLE figures the relative laser frequency was scanned from -30 GHz to 30 GHz (offset is \SI{327.112}{\tera\hertz})} 
   \label{fig:supp_ple_location_0}
\end{figure}

\begin{figure}[H]
    
  \includegraphics[width=1 \textwidth]{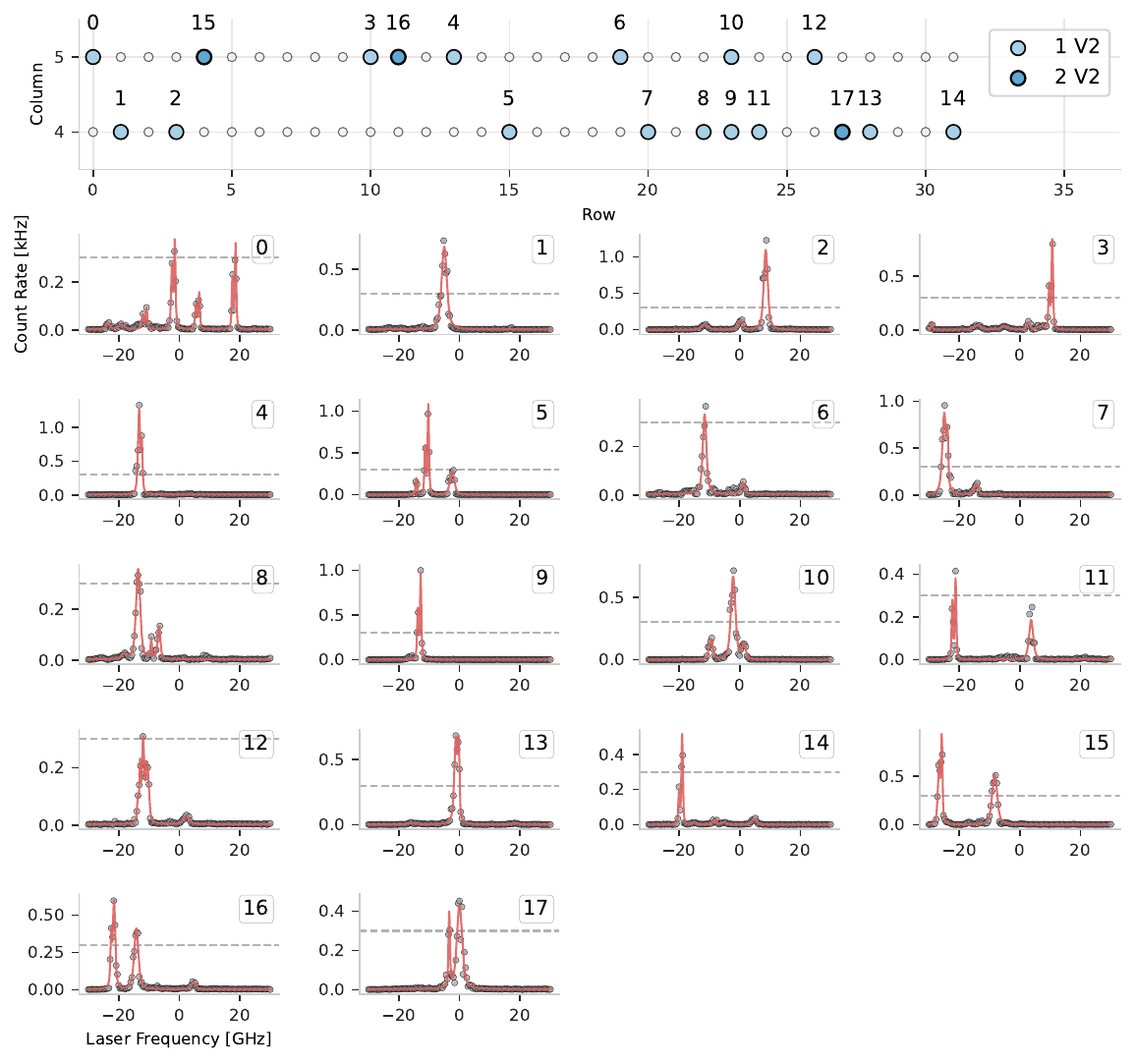}
  \caption{\textbf{Photoluminescence excitation of 64 nanopillars exposed to a single pulse of \SI{0.18}{\micro\joule}.} The white circles indicate a raster of nanopillars. The blue circles indicate nanopillars with V2 centers passing the \SI{0.3}{\kilo\hertz} threshold (grey dashed line), and their respective PLEs are shown below. Note that for all the PLE figures the relative laser frequency was scanned from -30 GHz to 30 GHz (offset is \SI{327.112}{\tera\hertz})} 
  \label{fig:supp_ple_location_018}
\end{figure}

\begin{figure}[H]
    
  \includegraphics[width=1 \textwidth]{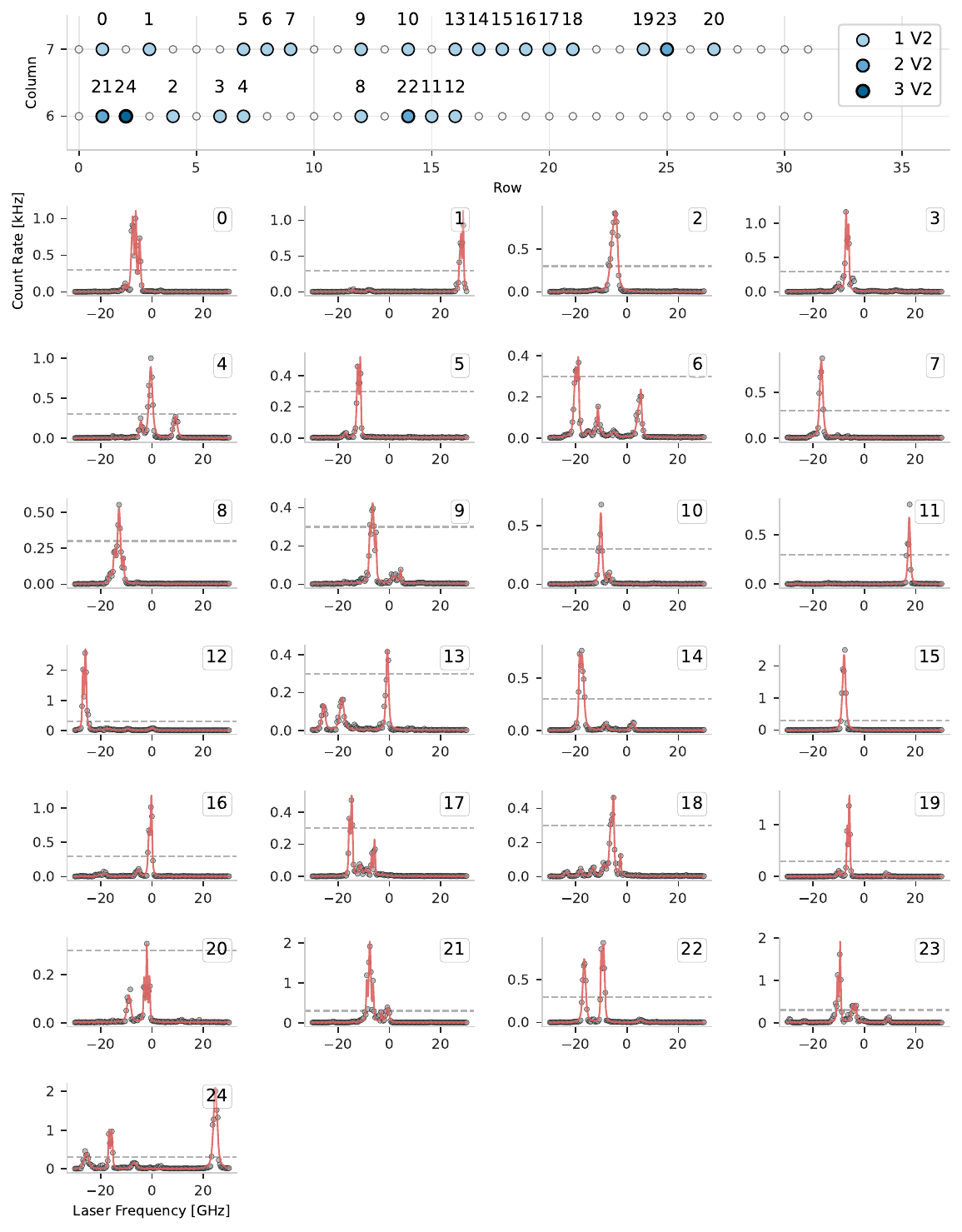}
  \caption{\textbf{Photoluminescence excitation of 64 nanopillars exposed to a single pulse of \SI{0.24}{\micro\joule}.} The white circles indicate a raster of nanopillars. The blue circles indicate nanopillars with V2 centers passing the \SI{0.3}{\kilo\hertz} threshold (grey dashed line), and their respective PLEs are shown below. Note that for all the PLE figures the relative laser frequency was scanned from -30 GHz to 30 GHz (offset is \SI{327.112}{\tera\hertz})} 
  \label{fig:supp_ple_location_024}
\end{figure}

To count the number of $\vtwo$ centers, we perform PLE measurements on each nanopillar according to the scheme indicated in the inset of \subfigref{fig:amound_defects}{a}.
The method is explained in the methods section of the main text. In \autoref{fig:supp_ple_location_0}, \autoref{fig:supp_ple_location_018}, \autoref{fig:supp_ple_location_024}, we show the spatial location and fitted PLE data for the unexposed,  \SI{0.18}{\micro\joule}, and \SI{0.24}{\micro\joule} nanopillars, respectively, which hosts `bright $\vtwo$ centers' passing the threshold. There seems to be no direct relationship between the spatial location of the nanopillars and the number of `bright $\vtwo$ centers'. All nanopillars are within \raisebox{0.4ex}{\tiny$\sim$}\SI{50}{\micro\meter} of the bondwire, and for each group of 64 nanopillars, the optical path was optimized on collection efficiency. Furthermore, the nanopillars shown in this section are all located within an area of \SI{250}{\micro\meter}$\times$\SI{20}{\micro\meter}.

\begin{figure}[H]
  \includegraphics[width=1 \textwidth]{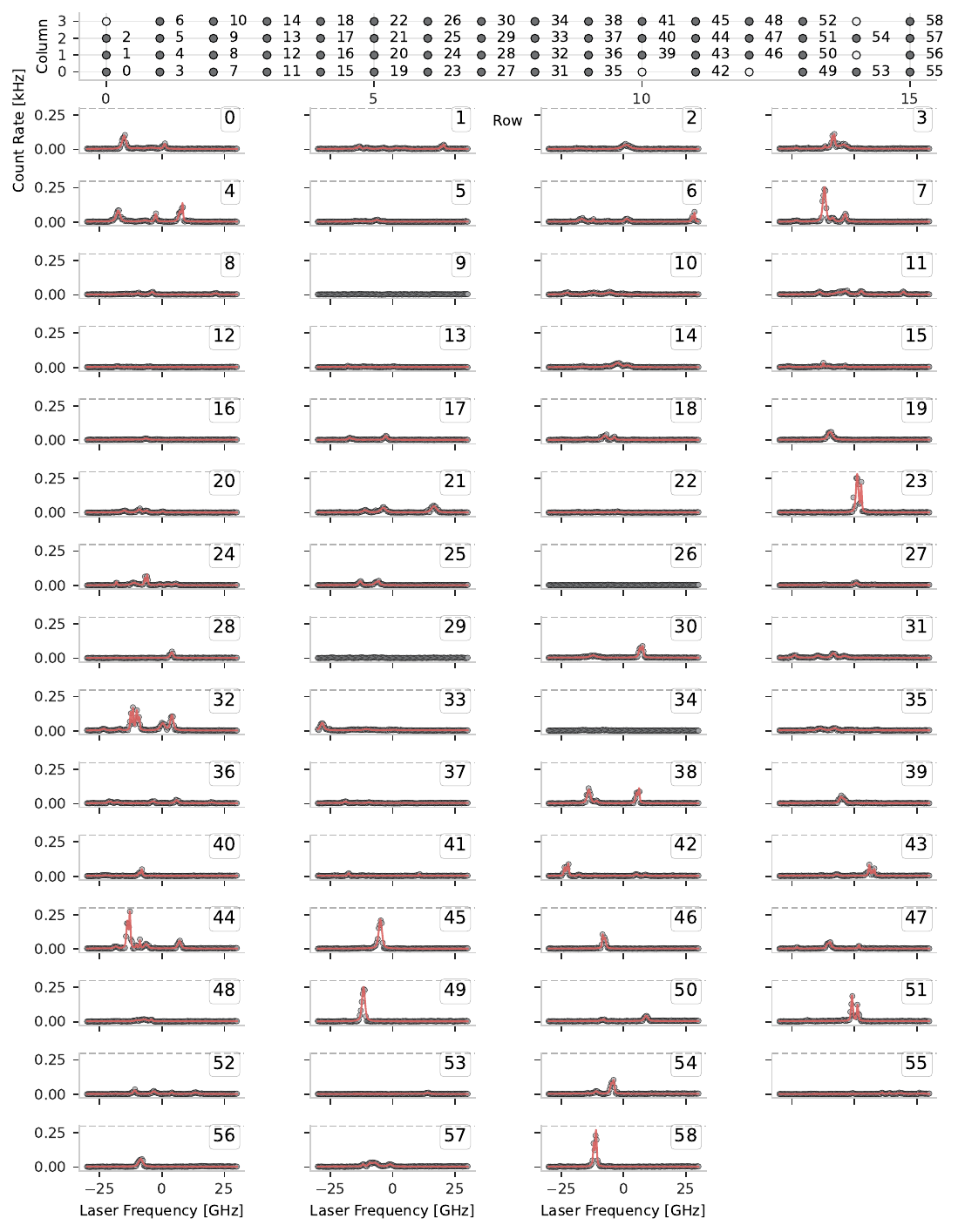}
  \caption{\textbf{Photoluminescence excitation of 64 unexposed nanopillars.} The grey circles indicate nanopillars with V2 centers not passing the \SI{0.3}{\kilo\hertz} threshold (grey dashed line), and their respective PLEs are shown below. Note that for all the PLE figures the relative laser frequency was scanned from -30 GHz to 30 GHz (offset is \SI{327.112}{\tera\hertz})} 
   \label{fig:supp_ple_location_0_below}
\end{figure}

\begin{figure}[H]
    
  \includegraphics[width=1 \textwidth]{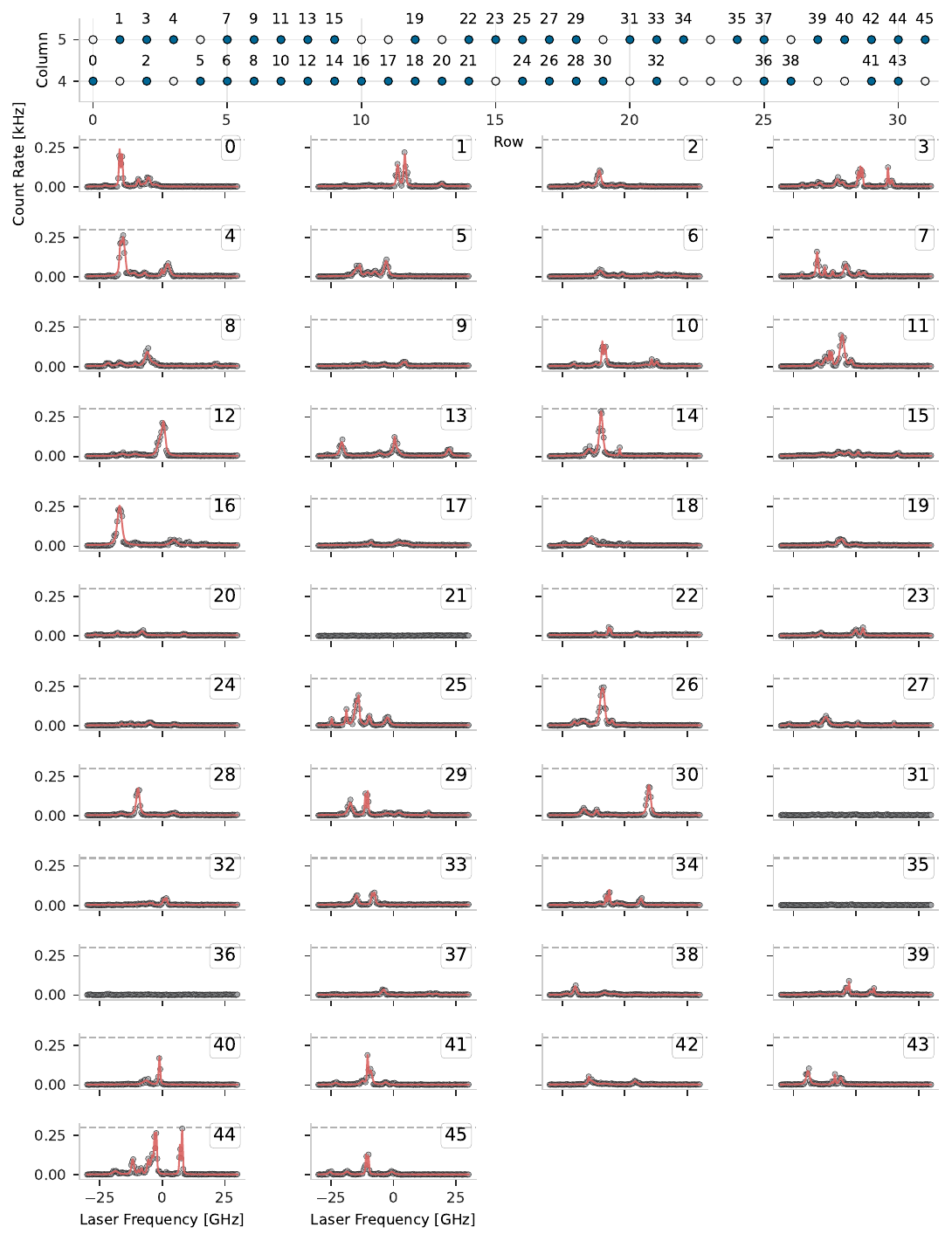}
  \caption{\textbf{Photoluminescence excitation of 64 nanopillars exposed to a single pulse of \SI{0.18}{\micro\joule}.} The white circles indicate a raster of nanopillars. The blue circles indicate pillars with V2 centers that did not pass the threshold, and their respective PLEs are shown below. Note that for all the PLE figures the relative laser frequency was scanned from -30 GHz to 30 GHz (offset is \SI{327.112}{\tera\hertz})} 
  \label{fig:supp_ple_location_018_below}
\end{figure}

\begin{figure}[H]
    
  \includegraphics[width=1 \textwidth]{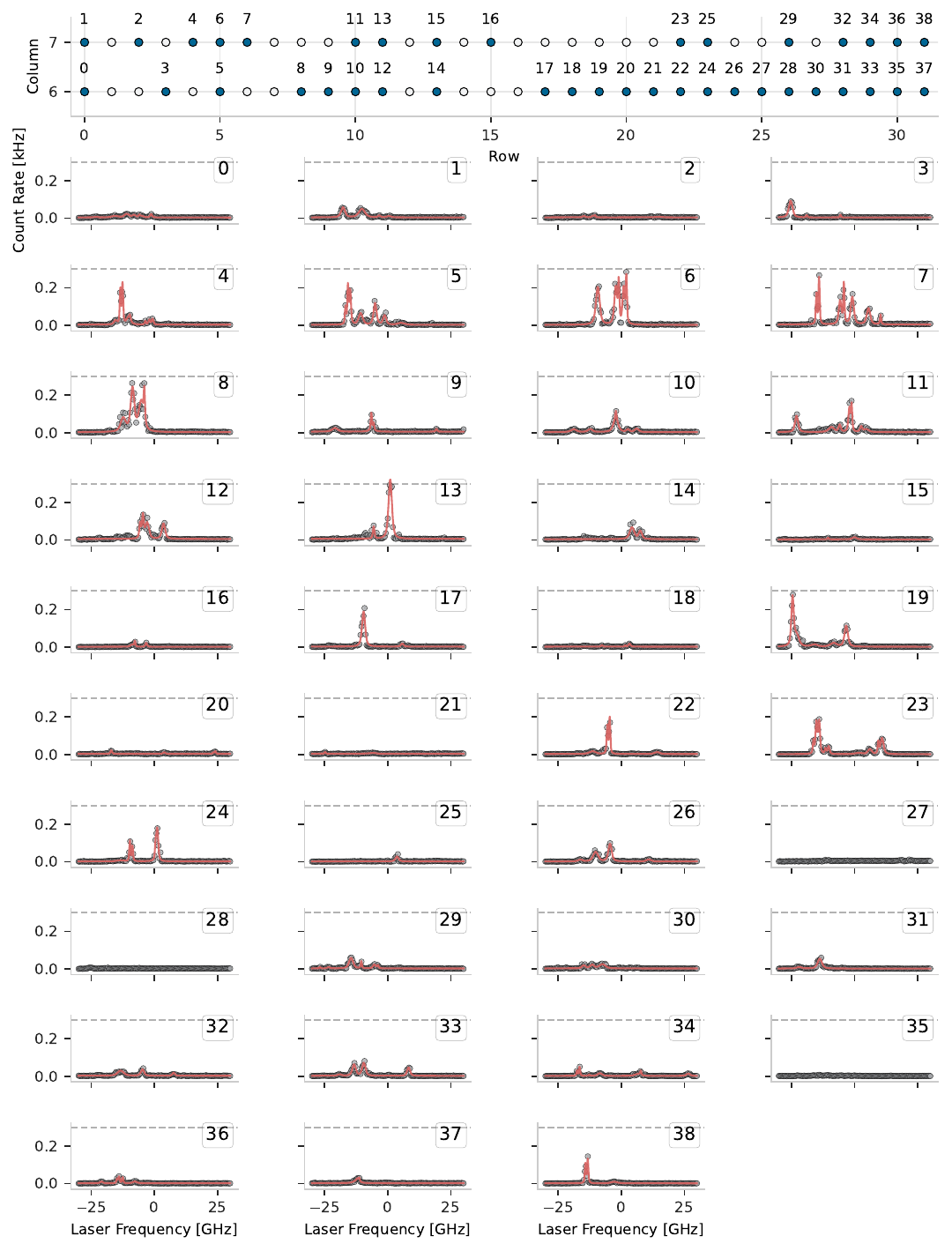}
  \caption{\textbf{Photoluminescence excitation of 64 nanopillars exposed to a single pulse of \SI{0.24}{\micro\joule}.} The white circles indicate a raster of nanopillars. The blue circles indicate pillars with V2 centers that did not pass the threshold, and their respective PLEs are shown below. Note that for all the PLE figures the relative laser frequency was scanned from -30 GHz to 30 GHz (offset is \SI{327.112}{\tera\hertz})} 
  \label{fig:supp_ple_location_024_below}
\end{figure}

All PLE spectra for nanopillars that contain $\vtwo$ centers and did not pass the threshold can be found in \autoref{fig:supp_ple_location_0_below}, \autoref{fig:supp_ple_location_018_below}, \autoref{fig:supp_ple_location_024_below}

\subsection{Spectral diffusion rate}
\label{subsec:supp_spectral_diffusion}
\begin{figure}[ht]
\centering
\label{fig:supp_diffusion_scheme}
  \includegraphics[width=0.3 \textwidth]{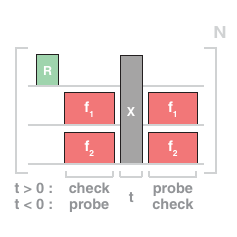}
  \caption{\textbf{Spectral diffusion measurement scheme.} Experimental sequence. A `check' block (\SI{2}{\milli \second}, \SI{20}{\nano \watt}) is followed by a system perturbation (marked `X'), which here consists either of turning on the NIR lasers or turning on the off-resonant laser. A second block (\SI{2}{\milli \second}, \SI{20}{\nano \watt}) probes whether the defect has diffused away, or has ionised (denoted `probe'). Data is post-selected by imposing a minimum-counts threshold ($T$), heralding the emitter on resonance in the first (second) block and computing the mean number of counts in the second (first) block, which encodes the emitter brightness at future (past) delay times $t$. To measure spectral diffusion under resonant excitation, we use two resonant lasers, each at $\SI{20}{\nano\watt}$. For spectral diffusion under off-resonant laser light, we use an off-resonant laser (\SI{785}{\nano\meter} with $\SI{10}{\micro\watt}$}).
\end{figure}

To measure the spectral diffusion rate, we employ the scheme shown in \ref{fig:supp_diffusion_scheme}. Here, we keep both resonant lasers on during the \textit{X-block}, and we use the no-recapture model to fit the data \cite{vandestolpeCheckprobeSpectroscopyLifetimelimited2025}. For off-resonant diffusion, we apply the off-resonant laser during the \textit{X-block} and fit the data to the diffusion-only model \cite{vandestolpeCheckprobeSpectroscopyLifetimelimited2025}.
The no-recapture model fits the data as:
\begin{equation}
    \Cmean(t)/\Co = \begin{cases}
    \left( 1 + \gdif \, t\, /\,\Gamma \right)^{-1} \,e^{-\gion t} , & \text{if $t > 0$}.\\
    \left( 1 - \gdif \, t\, /\,\Gamma \right)^{-1} , & \text{otherwise}.
    \end{cases}
\end{equation}

And the diffusion-only model fits the data as:
\begin{equation}
    \Cmean(t)/\Co = \begin{cases}
    \left( 1 + \gdif \, t\, /\,\Gamma \right)^{-1} , & \text{if $t > 0$}.\\
    \left( 1 - \gdif \, t\, /\,\Gamma \right)^{-1}, & \text{otherwise}.
    \end{cases}
\end{equation}

For both models, the spectral diffusion rate $\gdif$ depends on the linewidth of the specific $\vtwo$ center of interest. We assume that the linewidth is not lifetime-limited, and thus, if we want to extract the diffusion rate, we need to measure the homogeneous linewidth of each $\vtwo$. Since we did not measure the homogenous linewidth for each $\vtwo$ center, we take a range for $\gdif$ where:
\begin{equation}
    \gdif = \gamma_\mathrm{d,lifetime}\frac{\Gamma}{\Gamma_\mathrm{lifetime}}.
\end{equation}
Here, $\Gamma$ is the homogeneous linewidth of the emitter while $\gamma_\mathrm{d,lifetime}$ is the spectral diffusion rate assuming a lifetime-limited linewidth: $\Gamma_\mathrm{lifetime}$. We take a maximum diffusion rate for $\Gamma = 3\cdot\Gamma_\mathrm{lifetime}$ (see \ref{subsec:app_cp_ple_sweep}). 
\newpage
\subsection{Bayesian analysis for Check-Probe PLE}
\label{subsec:app_cp_ple_sweep}
\begin{figure}[ht]
\centering
  \includegraphics[width=0.25 \textwidth]{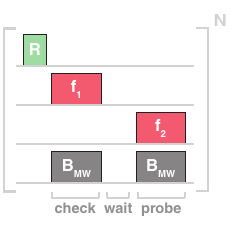}
  \caption{\textbf{Check-probe ple scheme.} An off-resonant laser pulse (\SI{10}{\micro\second} and\SI{10}{\micro\watt}) is applied followed by two resonant laser pulses (both \SI{2}{\milli\second} and \SI{20}{\nano\watt}) together with microwaves (\SI{127}{\mega\hertz}) to avoid spinpumping. The first resonant block acts as a check (effective initialisation) and the second resonant block acts as a probe (readout). The two blocks are separated by a waiting time of \SI{5}{\micro\second}.} 
   \label{fig:supp_cp_ple_scheme}
\end{figure}

We perform the check-probe PLE around the $A1$ transition to determine the homogeneous linewidths of several $\vtwo$ centers in the nanopillars of \subfigref{fig:diffusion_ple}{a,b}. The pulse sequence for the check-probe ple is shown in \autoref{fig:supp_cp_ple_scheme}. This type of ple allows to measure the homogeneous linewidths in systems where the rate of spectral diffusion is relatively high. Since the check and probe blocks are short (\SI{2}{\milli\second}) compared to the rate of spectral diffusion under resonant excitation.

\begin{figure}[ht]
\centering
  \includegraphics[width=1 \textwidth]{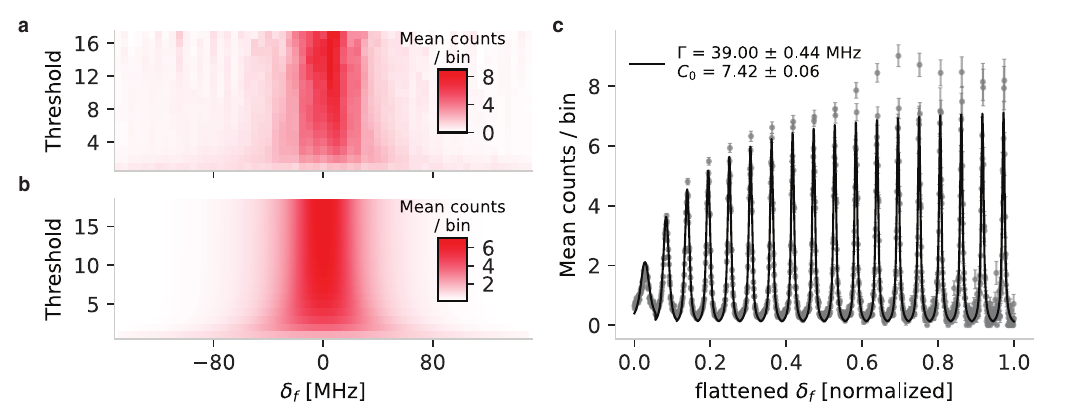}
  \caption{\textbf{Example of fitting check-probe signal.} \textbf{a)} Data obtained from pillar 21 (\autoref{fig:supp_ple_location_024}) where the threshold is swept from 1 to 17. \textbf{b)} Fitted convolution of \autoref{eq:convolution} to the data. \textbf{c)} Flattened data of a) and b) where the x axis is flattened and normalised. From the fit we extract $\Gamma = \SI{39.0\pm0.4}{\mega\hertz}$ and $C_0 = \SI{7.42\pm0.06}{}$ } 
   \label{fig:supp_cp_ple_sweep_example_fit}
\end{figure}

The probe spectrum obtained from this sequence typically contains some residual inhomogeneous broadening due to imperfect initialisation in the check step. To account for this, we analyse the data using the Bayesian framework introduced in van de Stolpe \emph{et al.}\cite{vandestolpeCheckprobeSpectroscopyLifetimelimited2025} (Supplementary Note 4). Within this model the mean number counts from the probe block at a frequency detuning $\mathbf{f}$ is expressed as the convolution of the spectral probability density (i.e. the residual inhomogeneous broadening) and the intrinsic emitter spectral response $\lambda(f)$:
\begin{equation} \label{eq:convolution}
    \Cmean(f) = P(f\,|\,\mc \geq T) \, * \, \lambda(f)  \, ,
\end{equation}
where the spectral probability density function is given by
\begin{equation} \label{eq:gamma_function}
    P(f\,|\,\mc \geq T) = \frac{1}{N_T} \left(1 - \Gamma_\mathrm{i}\left[  T, \lambda(f-f_1) \right]\right) \,,
\end{equation}
with $\lambda(f)$ the expected mean number of counts during obtained in a single check block when the emitter is at frequency $f$ and the laser at frequency $f_1$. The function $\Gamma_\mathrm{i}$ denotes the incomplete Gamma function
\begin{equation}
\Gamma_\mathrm{i}[a,z] = \frac{1}{\Gamma_\mathrm{c}[a]} \int_{z}^{\infty}t^{a-1}e^{-t}dt
\end{equation}
with $\Gamma_\mathrm{c}[a]$ the Euler-Gamma function. 
The emitters initrinsic spectral response is a Lorentzian described as
\begin{equation} \label{eq:lambda_f_L}
    \lambda_{\mathrm{L}}(f) = \Co \frac{\left(\frac{\Gamma}{2}\right)^2}{f^2 + \left(\frac{\Gamma}{2}\right)^2}.
\end{equation}
Because both terms in \autoref{eq:convolution} depend on $\lambda(f)$, the apparent spectral response varies with the set threshold $T$ from the check-block. And thus by increasing $T$ the amount of residual broadening becomes negligible. By fitting \autoref{eq:convolution} to the data where we sweep the threshold, we can extract the homogeneous linewidth $\Gamma$ and $C_0$. An example of the fit is shown in \autoref{fig:supp_cp_ple_sweep_example_fit}.

\begin{figure}[ht]
  \includegraphics[width=0.9 \textwidth]{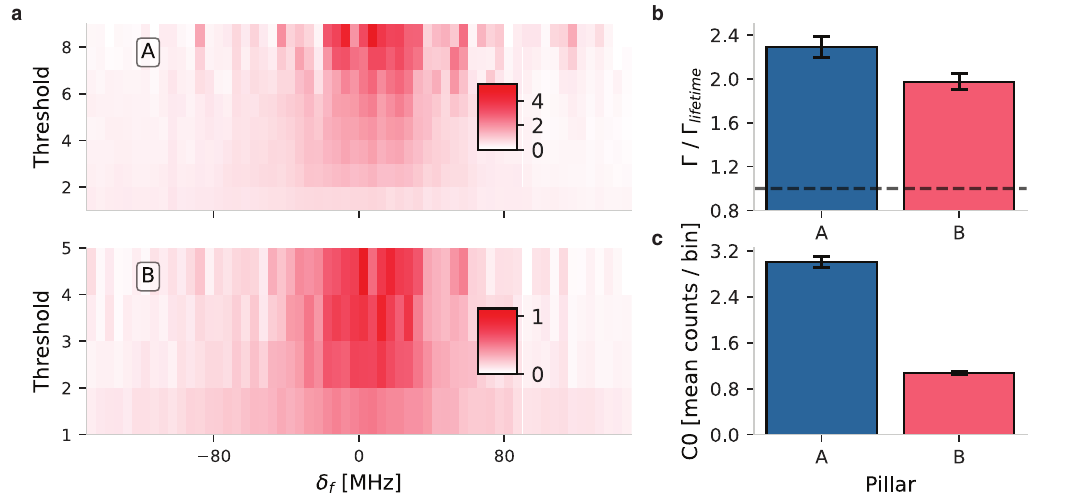}
  \caption{\textbf{$\Gamma$ for 2 natural $\vtwo$ centers.} \textbf{a)} Data of check-probe ple where threshold is swept for two $\vtwo$ centers. Which were not in \autoref{fig:supp_ple_location_0} but a different place. \textbf{b)} Extracted $\Gamma$ from fitting a) compared to $\Gamma_\mathrm{lifetime}$ for the same pillars. The dashed line indicates the lifetime limited $\Gamma$. \textbf{c)} Extracted $C_0$ from fitting a).  } 
   \label{fig:supp_cp_ple_sweep_0}
\end{figure}

\begin{figure}[ht]
  \includegraphics[width=0.865 \textwidth]{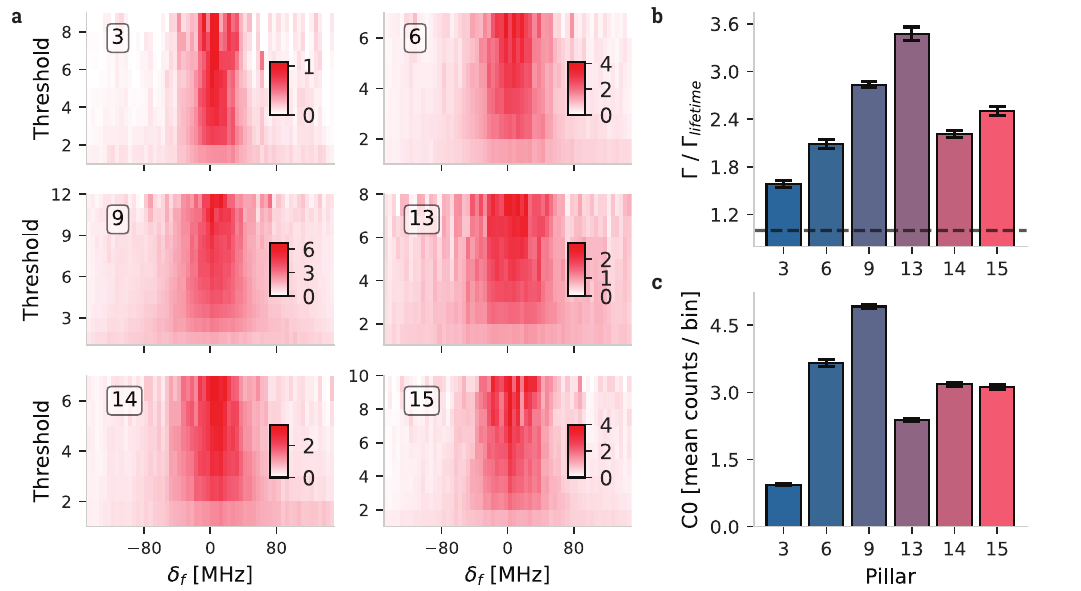}
  \caption{\textbf{$\Gamma$ for several laser-induced $\vtwo$ centers using UV pulse energy of \SI{0.18}{\micro\joule}.} \textbf{a)} Data of check-probe ple where threshold is swept for six $\vtwo$ centers. \textbf{b)} Extracted $\Gamma$ from fitting a) compared to $\Gamma_\mathrm{lifetime}$ for the same pillars. The dashed line indicates the lifetime limited $\Gamma$.  \textbf{c)} Extracted $C_0$ from fitting a).}
   \label{fig:supp_cp_ple_sweep_018}
\end{figure}

\begin{figure}[ht]
  \includegraphics[width=0.865 \textwidth]{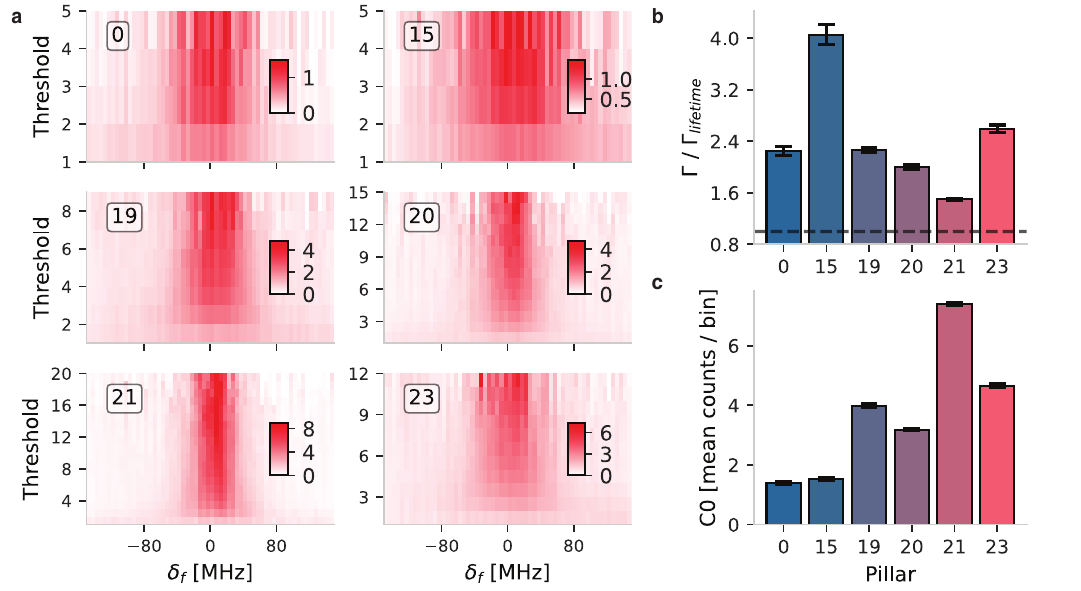}
  \caption{\textbf{$\Gamma$ for several laser-induced $\vtwo$ centers using UV pulse energy of \SI{0.24}{\micro\joule}.} \textbf{a)} Data of check-probe ple where threshold is swept for six $\vtwo$ centers. \textbf{b)} Extracted $\Gamma$ from fitting a) compared to $\Gamma_\mathrm{lifetime}$ for the same pillars. The dashed line indicates the lifetime limited $\Gamma$. \textbf{c)} Extracted $C_0$ from fitting a).}
   \label{fig:supp_cp_ple_sweep_024}
\end{figure}

Threshold sweeps were performed for for several natural $\vtwo$ centers (\autoref{fig:supp_cp_ple_sweep_0}) and laser-induced $\vtwo$ centers (single \SI{0.18}{\micro\joule} pulse in \autoref{fig:supp_cp_ple_sweep_018} and single \SI{0.24}{\micro\joule} pulse in \autoref{fig:supp_cp_ple_sweep_024}). The resulting linewidths are compared to the lifetime-limited value of \raisebox{0.4ex}{\tiny$\sim$}$\SI{26}{\mega\hertz}$ for the $A1$  transition. Note that due to the initialisation in the check-block with a single laser and a microwaves we also have a small probability to initialise in the $\pm\frac{3}{2}$ subspace. This introduces a small probability of measuring the A2 transition which has a longer lifetime and can therefore result in an lower measured $\Gamma$.
 
For measured two $\vtwo$ natural $\vtwo$ centers (different ones from \ref{subsec:app_location_defects} and observe linewidths \raisebox{0.4ex}{\tiny$\sim$}$2 \cdot \Gamma_\mathrm{lifetime}$. This indicates that we still have some residual inhomogeneous broadening and/or power broadening effects. Since the measurements were highly automatised, the count rate was not optimised perfectly before each measurement and can influence the spectral density function (thresholding) significantly. 
Laser-induced $\vtwo$ centers exhibit comparable $\frac{\Gamma}{\Gamma_\mathrm{lifetime}}$ and thus show that the linewidth is not significantly broadened for laser-induced $\vtwo$ centers and close to the lifetime limit. A more precise determination of the homogeneous linewidth, including Rabi-induced broadening, could be obtained with a low magnetic field (\raisebox{0.4ex}{\tiny$\sim$}\SI{0}{\gauss}) and high microwave power, so that Landau-Zener-St\"uckelberg interference becomes observable. These experiments lie beyond the scope of this study.

\clearpage
\subsection{Optical properties of other laser-induced V2 center}
\label{subsec:app_diffusion_spin_v2}
\begin{figure}[ht]
\centering
  \includegraphics[width=0.9 \textwidth]{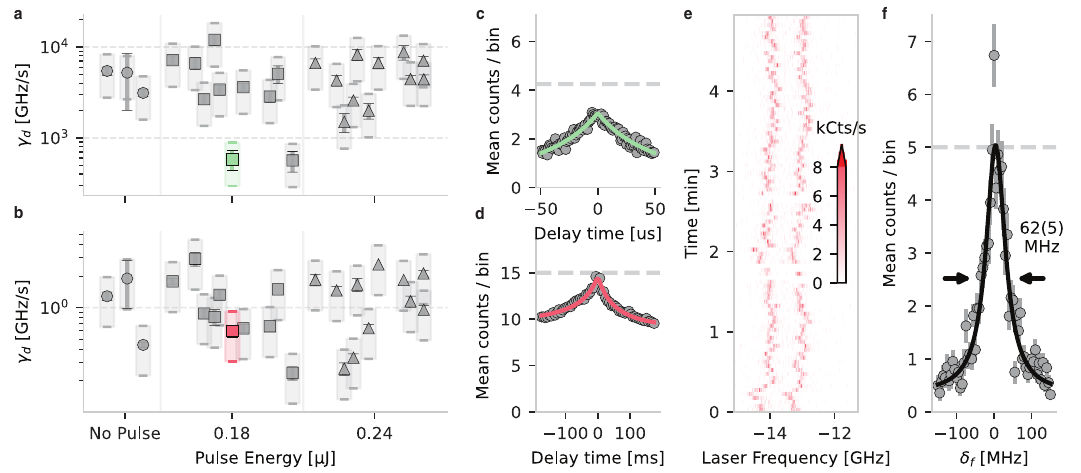}
  \caption{\textbf{Emitter optical properties and laser-induced charge dynamics of $\vtwo$ used for spin coherence measurements.} \textbf{a)} Fitted spectral diffusion constant under offresonant excitation for natural $\vtwo$ centers (circles) and laser-induced $\vtwo$ centers (squares and triangles, \SI{0.18}{\micro\joule} and \SI{0.24}{\micro\joule} respectively). The error bars indicate error in fit, the shaded region indicates the $\gdif$ interval for $1\leq\Gamma/\Gamma_\mathrm{lifetime}\leq3$ (see \ref{subsec:supp_spectral_diffusion}). $\gdif$ is extracted using methodology described in van de Stolpe \emph{et al.} \cite{vandestolpeCheckprobeSpectroscopyLifetimelimited2025}. $\gdif$ of the red square indicates the same $\vtwo$ center as the green square in b) and data in c), d), e) and f). \textbf{b)} Fitted spectral diffusion constants under resonant excitation for the same $\vtwo$ centers as in a). \textbf{c)} Data and fit to extract $\gamma_\mathrm{diff}$ under off-resonant excitation. \textbf{d)} Data and fit to extract $\gamma_\mathrm{diff}$ under resonant excitation. \textbf{e)} Scanning laser PLE over 5 minutes. The laser is on for a total of \SI{334}{\milli\second} per scan. We conditionally apply an off-resonant pulse if a single scan gives less than 5 counts. This is done to counter ionisation (several scans show no counts or only a single optical transition). \textbf{f} Check-Probe PLE (see van de Stolpe \emph{et al.}\cite{vandestolpeCheckprobeSpectroscopyLifetimelimited2025} for methodology) indicating a near-lifetime limited optical linewidth, performed at a magnetic field of \raisebox{0.4ex}{\tiny$\sim$}\SI{20}{\gauss}} 
  \label{fig:app_diffusion_spin_v2}
\end{figure}

For the measurements on the spin properties, we use one of the brightest V2 centers (pillar 9 \autoref{fig:supp_ple_location_018}). This V2 center also has a relatively low spectral diffusion rate under off-resonant excitation. In \autoref{fig:app_diffusion_spin_v2} we show the diffusion rates under off-resonant and resonant excitation, together with a scanning PLE and check-probe PLE. We observe a slightly larger linewidth (\raisebox{0.4ex}{\tiny$\sim$}$2\cdot\Gamma_\mathrm{lifetime}$ see \ref{subsec:app_cp_ple_sweep}) and slightly faster spectral diffusion rate compared to the V2 center in the main text.
To confirm we are solely addressing a single $\vtwo$ center we perform a $g^2$-correlation measurement. We apply limited off-resonant power ($\SI{1}{\micro\watt}$ to avoid ionisation and reduce background counts) and two resonant lasers at the frequency of the A1 and A2 transition (both $\SI{20}{\nano\watt}$ and collect data for $\SI{20}{\minute}$)
\begin{figure}[ht]
\centering
  \includegraphics[width=0.42\textwidth]{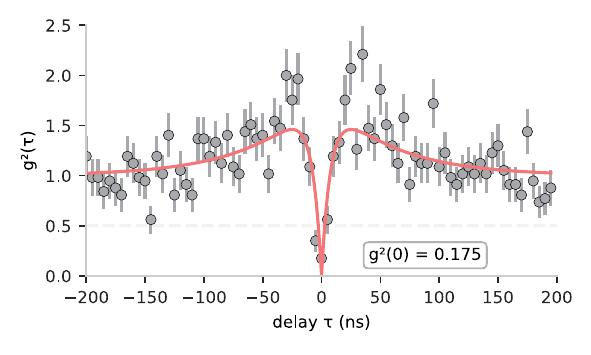}
  \caption{\textbf{$g^2$-correlation measurement under resonant excitation.} We apply limited off-resonant power ($\SI{1}{\micro\watt}$ to avoid ionisation and reduce background counts) and two resonant lasers at the frequency of the A1 and A2 transition (both $\SI{20}{\nano\watt}$ and collect data for $\SI{20}{\minute}$). The data for $g^2(0) = 0.175$ indicates that this is indeed a single $\vtwo$ center that we can address resonantly.}
  \label{fig:app_g2_correlation}
\end{figure}
\subsection{Electron spin measurements}
\label{subsec:app_spin_stuff}
\begin{figure}[ht]
  \includegraphics[width=10 cm]{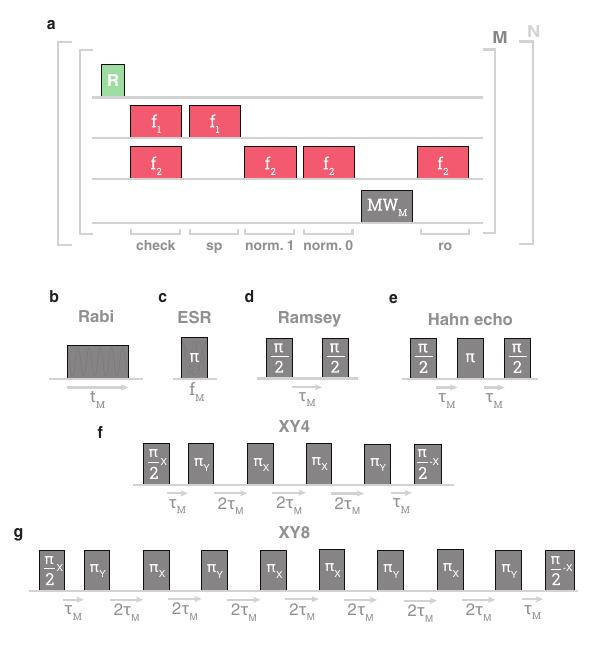}
  \caption{ \textbf{Pulse schemes for spin measurements.} \textbf{a)} A framework of pulse sequences shared by Rabi, electron spin resonance (ESR), Ramsey, Hahn echo, and dynamical decoupling (DD) measurements is reported in the main text. R represents a \SI{785}{\nano\meter} repump laser, $\text{f}_1$ a laser resonant with the A1 transition, $\text{f}_2$ a laser resonant with the A2 transition, and $\text{MW}_\text{M}$ some parametrized microwave sequence that depends on the specific measurement and the parameter $M$. In each measurement, we run the entire laser and microwave sequence $M$ times, sweeping the parameter $M$ for the MW sequence, and we repeat the sweep $N$ times to mitigate errors such as shot noise. \textbf{b)} In Rabi measurements, we fix the frequency and power of MW driving and sweep its duration $t_M$. \textbf{c)} In ESR, we send a MW $\pi$-pulse for fixed duration and power and sweep its frequency $f_M$. \textbf{d)} In Ramsey measurements, we sweep MW sequences of the form $(\frac{\pi}{2}-\tau_M-\frac{\pi}{2})$, where $\pi/2$ represents a MW $\pi/2$-pulse with fixed duration, frequency and power and $\tau_M$ represents the variable interpulse delay. \textbf{e)} In Hahn echo measurements, we sweep MW sequences of the form $(\frac{\pi}{2}-\tau_M-\pi-\tau_M-\frac{\pi}{2})$, with $\frac{\pi}{2}$, $\pi$ and $\tau_M$ similarly defined. \textbf{f) g)} In DD measurements, we run XY4 and XY8 sequences and sweep the interpulse delay $2\tau_M$. $\frac{\pi}{2}_X$ and $\frac{\pi}{2}_{-X}$ represent MW $\frac{\pi}{2}$ pulses with a phase of 0° and 180° respectively, and $\pi_X$ and $\pi_Y$ represent MW $\pi$ pulses with a phase of 0° and 90° respectively,} 
  \label{fig:supp_spin_sequence}
\end{figure}

\begin{figure}[ht]
  \includegraphics[width=1 \textwidth]{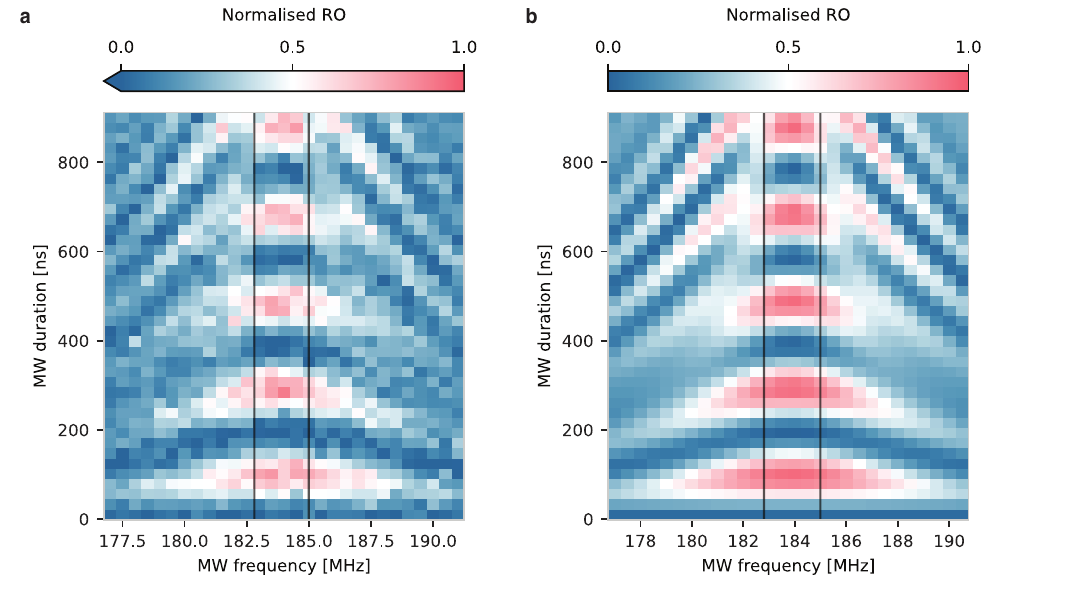}
  \caption{\textbf{Measured and simulated Rabi Chevron patterns of Pillar 9.} Solid dark lines in each panel correspond to resonant frequencies fitted in the DESR spectrum (\subfigref{fig:electron_spin}{b} in main text). \textbf{a)} Measured data, also displayed in \subfigref{fig:electron_spin}{a}. \textbf{b)} Simulated with the hyperfine coupling extracted from the DESR spectrum.} 
\end{figure}

To investigate the electron spin properties of laser-induced $\vtwo$ centers, we perform Rabi oscillation, electron spin resonance (ESR), Ramsey, Hahn echo and dynamical decoupling (DD) measurements on pillars 9 and 14(\autoref{fig:supp_ple_location_018}). The pulse sequences for all experiments are summarized in \autoref{fig:supp_spin_sequence}. We begin by resetting the spin environment of the $\vtwo$ center with a \SI{10}{\micro\second} \SI{785}{\nano\meter} laser pulse in the \textbf{repump} block. Then, we verify that the $\vtwo$ center is on resonance by illuminating both resonant lasers for \SI{150}{\micro\second} in the \textbf{check} block. Subsequently, we populate the $m_s=±\frac{3}{2}$ subspace by exciting the A1 transition for \SI{60}{\micro\second} in the spin-pump (\textbf{sp}) block. This is followed by two identical normalization blocks (\textbf{norm.1} and \textbf{norm.0}), each exciting the A2 transition for \SI{60}{\micro\second}. Photon counts are recorded in both blocks. We then apply appropriate microwave pulse sequences to run different measurements in the $\textbf{MW}_\textbf{M}$ block, before reading out the population in $m_s=±\frac{3}{2}$ subspace with a \SI{60}{\micro\second} A2 laser pulse in the readout (\textbf{ro}) block. A wait time of \SI{10}{\micro\second} is inserted between adjacent blocks.

We apply a threshold on the photon count rates recorded during the \textbf{check} block. For those repetitions that pass the threshold, we calculate the mean of photon counts during \textbf{norm.1}, \textbf{norm.0} and \textbf{ro} blocks, denoted as $A$, $B$ and $C$ respectively, as well as the unbiased estimators of standard deviations $\sigma_A$, $\sigma_B$ and $\sigma_C$. The normalized readout $R$ is then calculated as
\begin{equation}
    R = \frac{C-B}{A-B},
\end{equation}
and the uncertainty $\sigma_R$ is propagated as
\begin{equation}
    \sigma_R = \sqrt{\frac{(C-B)^2}{(A-B)^4}\sigma^2_A + \frac{(A-C)^2}{(A-B)^4}\sigma^2_B + \frac{1}{(A-B)^2}\sigma^2_C}.
\end{equation}

In ESR measurements, we fit the normalized readout $R$ to 2 Gaussian peaks with a shared full-width at half-maximum (FWHM). We then extract a \Ttwostar using
\begin{equation}
    T_{2,\text{DESR}}^* = \frac{2\sqrt{\ln2}}{\pi\cdot\text{FWHM}}.
\end{equation}

In detuned Ramsey measurements, we fit the normalized readout $R$ to the following function:
\begin{equation}
    R = b + \mathrm{e}^{-(\tau/T^*_2)^2} \sum_iA_i\cos\bigg[\bigg(f_c+\frac{(-1)^i}{2}f_\text{HF}\bigg)\tau+\varphi_i\bigg],
\end{equation}
where $\tau$ is the interpulse delay, $i=0,1$ labels the peaks fitted in the DESR spectrum, $A_i$ and $\varphi_i$ are the amplitude and phase of the oscillation due to detuning from each DESR peak, $f_c$ is the center or average frequency of all oscillations, $f_\text{HF}$ is the hyperfine coupling of the nuclear spin, $b$ is a global offset and $\Ttwostar$ is the extracted dephasing time. For the fit in the main text (\subfigref{fig:electron_spin}{c}) we extract $f_c = \SI{2.13\pm0.03}{\mega\hertz}$.

In Hahn echo measurements, we fit the normalized readout $R$ to a stretched decay function:
\begin{equation}
    R = b + A\mathrm{e}^{-(t/T_2)^n},
\end{equation}
where $t=2\tau$ is the total time the V2 electron spin is allowed to dephase on the $xy$-plane of the Bloch sphere, $\tau$ is the interpulse delay, $A$ is the amplitude of the decay, $b$ is a global offset, $T_2$ is the extracted coherence time and $n$ is the decay exponent.

In DD experiments, we fit the same stretched decay function. Here, we have $t=2N\tau$ where $N$ is the number of decoupling $\pi$ pulses and $2\tau$ is the wait time between two successive $\pi$ pulses.

\subsection{Spin coherence of V2 center in pillar 14 at low magnetic field}
\label{subsec:app_spin_4_32}
\begin{figure}[ht]
  \includegraphics[width=1 \textwidth]{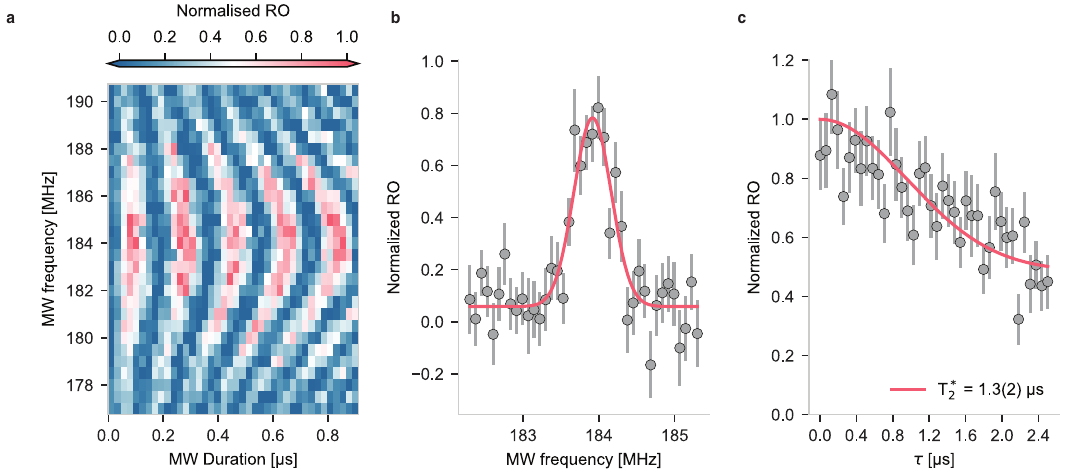}
  \caption{\textbf{Spin measurements at low field of $\vtwo$ center in pillar 14:} \textbf{a)} Rabi chevron pattern. Details on the measurement sequence and normalisation procedure are elaborated in \ref{subsec:app_spin_stuff}. \textbf{b)} Electron-spin-resonance of $\vtwo$ center revealing no strongly coupled nuclear spin. \textbf{c)} Resonant Ramsey measurement on the same $\vtwo$ center with $f_{MW} = \SI{183.8}{\mega\hertz}$. Red line indicates a fit to a Gaussian decay with $\Ttwostar = \SI{1.3\pm0.2}{\micro\second}$. }
  \label{fig:supp_spin_4_32}
\end{figure}

We also characterised the spin properties of the highlighted $\vtwo$ center in \autoref{fig:diffusion_ple} (pillar 14 \autoref{fig:supp_ple_location_018}). For this $\vtwo$ center, we could not find a (resolvable) strongly coupled nuclear spin. See \subfigref{fig:supp_spin_4_32}{a,b} for the Rabi chevron pattern and ESR measurement. We find a similar spin coherence time $\Ttwostar = \SI{1.3\pm0.2}{\micro\second}$ (see \subfigref{fig:supp_spin_4_32}{c}) as the $\vtwo$ center shown in \autoref{fig:electron_spin}.
\subsection{Optical setups}
\label{subsec:app_optical_setup}
For this work, we used 2 confocal setups, one at both ambient pressure and temperature, and one cryogenic setup (4K).

\subsubsection{Ambient-temperature setup}
We couple an \SI{785}{\nano \meter} off-resonant laser (Cobolt 06-MLD785) to free space using a zoom fiber collimator (Thorlabs, ZC618APC).
Free Space: The collimated beam passes through a variable neutral density filter (ND, Thorlabs NDC-50C-4-B), after which a $\lambda/2 − \lambda/4$ waveplate combination allows for polarisation control. The excitation path and detection path are separated with a $\SI{925}{\nano\meter}$ dichroic beamsplitter (Semrock FF925-Di01-25x36). We use a nanosecond-pulsed nitrogen laser (LaserTechnik
Berlin MNL 100 ) which delivers $\SI{3}{\nano\second}$ pulses at $\SI{337.1}{\nano\meter}$. We control the energy of the UV pulses using a neutral density filter on a servo motor (coarse) and the laser's voltage (fine). We measure the pulse energy just before the objective using a $\micro$-Joule Meter (PEM 250, LaserTechnik Berlin). Using a 355 dichroic beamsplitter (Semrock, Di01-R355-25x36) we align the UV laser to the sample. A flip mirror and 50/50 pellicle beamsplitter (Thorlabs, BP150) enable imaging of the sample with a visible LED (MCWHL6-C2) and a CCD-camera (ClearView Imaging, BFS-U3-16S2M-CS). Two different objectives are used to focus the laser onto the sample: a 60X, 0.7 NA (Olympus LUCPLFLN60X) and a 100×, 0.9 NA objective ([137]). The 60X, 0.7 NA objective is employed for laser writing because of its compatibility with UV wavelengths. For room-temperature defect characterization, we use the 100X, 0.9 NA objective, which offers improved photon collection, but does not support UV transmission. The objective can be moved using a configuration of 3 piezo-electric stages (PI Q545.140). 
Collected fluorescence passes through both dichroic beamsplitters and is filtered with an 830 nm long-pass filter (Semrock, BLP01-830R-25) and collected in a fiber. We perform the 2D-PL measurements without a 50/50 beamsplitter, but for the saturation measurements, we used a 50/50 beamsplitter as we also performed $g^2$-correlation measurements. We use two single-photon superconducting nanowire
single-photon detectors (SNSPDs, Single Quantum) with over 90\% detection efficiency at 917 nm.
The coarse time scheduling (1 μs resolution) of the experiments is managed by a microcontroller
(ADwin Pro II)

\begin{figure}[ht]
  \includegraphics[width=1 \textwidth]{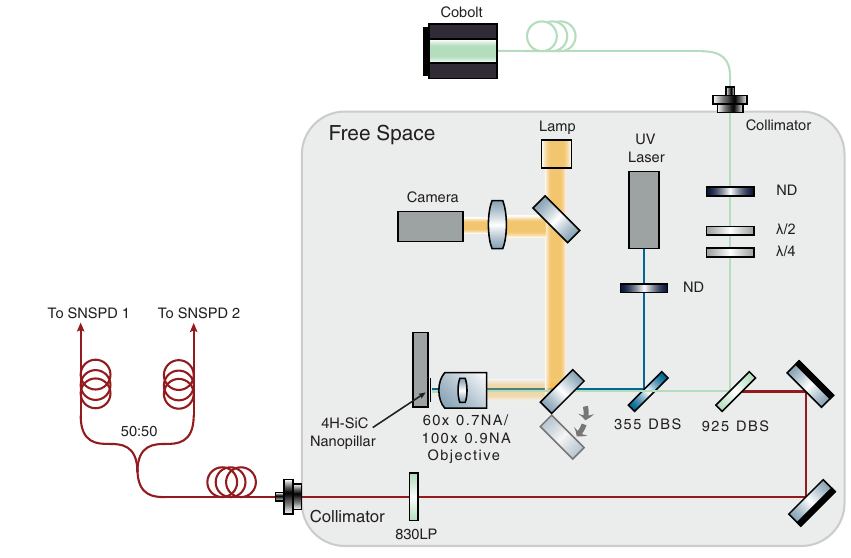}
  \caption{ \textbf{Room-temperature confocal optical setup.} Details in text above the figure.} 
\end{figure}
\subsubsection{Low-temperature optical setup}
The low-temperature optical setup is divided into two parts, in-fiber (left) and free-space (right). The electronics are not depicted in the figure.\\
\textbf{In Fiber:} Two NIR lasers (\SI{916}{\nano \meter}, Toptica DL Pro and the Spectra-Physics Velocity TLB-6718-P, are frequency-locked to a wavemeter (HF-Angstrom WS/U-10U), using a 99/1 beamsplitter. Their optical power is modulated by acousto-optic modulators (AOM, G\&H SF05958). The power of the \SI{785}{\nano \meter} repump laser (Cobolt 06-MLD785) is directly controlled via analog modulation. A wavelength division multiplexer (WDM, OZ Optics) combines the \SI{785}{\nano \meter} repump and \SI{916}{\nano \meter} NIR laser light, after which the light is coupled out to free space using a zoom fiber collimator (Thorlabs, ZC618APC). \\
\textbf{Free Space:} The collimated beam passes through a variable neutral density filter (ND, Thorlabs NDC-50C-4-B), after which a shortpass filter at an angle (Semrock, FF01-945/SP-25) is used to remove any residual noise from the NIR lasers. A $\frac{\lambda}{2}-\frac{\lambda}{4}$ waveplate combination allows for polarisation control. The excitation and detection paths are separated by a broadband 90:10 beamsplitter (Thorlabs, BS041). We use a fast-steering mirror (Newport, FSM-300-02) and a 4f system to scan the lasers over the sample. A flip mirror and 50/50 pellicle beamsplitter (Thorlabs, BP150) enable imaging of the sample with a visible LED (MCWHL6-C2) and a CCD-camera (ClearView Imaging, BFS-U3-16S2M-CS).

A 0.9 NA microscope objective (Olympus, MPLFLN 100X) is used to focus excitation light onto the nanopillars and to collect fluorescence. The objective is kept at room temperature and under vacuum, and is moved (coarse) using a configuration of 3 piezo-electric stages (PI Q545.140). The sample is cooled down to \SI{4}{\kelvin} in a closed-cycle cryostat (Montana Instruments S100), while a heat shield kept at \SI{30}{\kelvin} limits thermal radiation from the objective. 

Collected fluorescence passes through a 90/10 beamsplitter, after which it can be routed either to a spectrometer (Princeton Instruments IsoPlane 81), filtered with an \SI{830}{\nano \meter} long-pass filter (Semrock, BLP01-830R-25), or to the single photon superconducting nanowire single-photon detectors (SNSPDs, Single Quantum) with over 90\% detection efficiency at 917 nm. Next to a \SI{830}{\nano \meter} long pass filter, an additional long-pass filter (Semrock, FF01-937/LP-25), placed at an angle, is used to filter out reflected light originating from the NIR lasers (916 nm).\\
\textbf{Electronics:}
Microwave pulses are generated with an arbitrary waveform generator (Zurich Instruments, HDAWG8) and subsequently amplified (Mini-circuits LZY-22+). A bondwire is spanned across the sample to deliver the MW radiation close to the sample surface (\raisebox{0.4ex}{\tiny$\sim$}$\SI{25}{\micro \meter}$). The coarse-time scheduling ($\SI{1}{\micro \second}$ resolution) of the experiments is handled by a microcontroller (ADwin Pro II).
\begin{figure}[ht]
  \includegraphics[width=1 \textwidth]{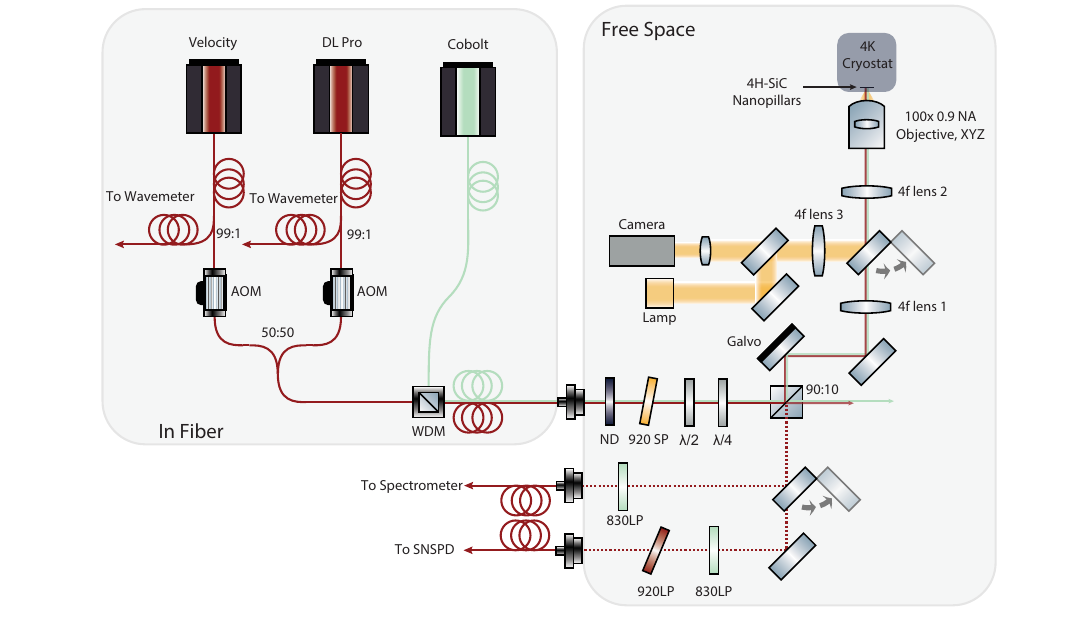}
  \caption{ \textbf{Low-temperature confocal optical setup.} Details in text above the figure.} 
\end{figure}

\clearpage

\end{document}